\documentclass[twocolumn]{aastex631}
\usepackage{amsmath}
\usepackage{ulem}

\shorttitle{NIR DIBs in $Y$ and $J$ bands}
\shortauthors{Hamano et al.}

\begin{document}

\title{Survey of near-infrared diffuse interstellar bands in $Y$ and $J$ bands. 

I. Newly identified bands}

\correspondingauthor{Satoshi Hamano}

\author[0000-0002-6505-3395]{Satoshi Hamano}
\affiliation{National Astronomical Observatory of Japan, 2-21-1 Osawa, Mitaka, Tokyo 181--8588, Japan}
\affiliation{Laboratory of Infrared High-resolution Spectroscopy (LiH), Koyama Astronomical Observatory, Kyoto Sangyo University, Motoyama, Kamigamo, Kita-ku, Kyoto 603--8555, Japan}
\email{satoshi.hamano@nao.ac.jp}

\author{Naoto Kobayashi}
\affiliation{Kiso Observatory, Institute of Astronomy, School of Science,The University of Tokyo, 10762-30 Mitake, Kiso-machi, Kiso-gun, Nagano, 397--0101, Japan}

\author[0000-0003-2011-9159]{Hideyo Kawakita}
\affiliation{Laboratory of Infrared High-resolution spectroscopy(LiH), Koyama Astronomical Observatory, Kyoto Sangyo University, Motoyama, Kamigamo, Kita-ku, Kyoto 603--8555, Japan}
\affiliation{Department of Physics, Faculty of Sciences, Kyoto Sangyo University, Motoyama, Kamigamo, Kita-ku, Kyoto 603--8555, Japan}

\author{Keiichi Takenaka}
\affiliation{Laboratory of Infrared High-resolution spectroscopy(LiH), Koyama Astronomical Observatory, Kyoto Sangyo University, Motoyama, Kamigamo, Kita-ku, Kyoto 603--8555, Japan}
\affiliation{Department of Physics, Faculty of Sciences, Kyoto Sangyo University, Motoyama, Kamigamo, Kita-ku, Kyoto 603--8555, Japan}

\author[0000-0003-2380-8582]{Yuji Ikeda}
\affiliation{Laboratory of Infrared High-resolution spectroscopy(LiH), Koyama Astronomical Observatory, Kyoto Sangyo University, Motoyama, Kamigamo, Kita-ku, Kyoto 603--8555, Japan}
\affiliation{Photocoding, 460-102 Iwakura-Nakamachi, Sakyo-ku, Kyoto, 606--0025, Japan}

\author{Noriyuki Matsunaga}
\affiliation{Department of Astronomy, Graduate School of Science, University of Tokyo, Bunkyo-ku, Tokyo 113--0033, Japan}
\affiliation{Laboratory of Infrared High-resolution spectroscopy(LiH), Koyama Astronomical Observatory, Kyoto Sangyo University, Motoyama, Kamigamo, Kita-ku, Kyoto 603--8555, Japan}

\author{Sohei Kondo}
\affiliation{Kiso Observatory, Institute of Astronomy, School of Science,The University of Tokyo, 10762-30 Mitake, Kiso-machi, Kiso-gun, Nagano, 397--0101, Japan}
\affiliation{Laboratory of Infrared High-resolution spectroscopy(LiH), Koyama Astronomical Observatory, Kyoto Sangyo University, Motoyama, Kamigamo, Kita-ku, Kyoto 603--8555, Japan}

\author[0000-0001-6401-723X]{Hiroaki Sameshima}
\affiliation{Institute of Astronomy, School of Science, University of Tokyo,2-21-1 Osawa, Mitaka, Tokyo 181--0015, Japan}
\affiliation{Laboratory of Infrared High-resolution spectroscopy(LiH), Koyama Astronomical Observatory, Kyoto Sangyo University, Motoyama, Kamigamo, Kita-ku, Kyoto 603--8555, Japan}

\author{Kei Fukue}
\affiliation{Laboratory of Infrared High-resolution spectroscopy(LiH), Koyama Astronomical Observatory, Kyoto Sangyo University, Motoyama, Kamigamo, Kita-ku, Kyoto 603--8555, Japan}
\affiliation{Education Center for Medicine and Nursing, Shiga University of Medical Science, Seta Tsukinowa-cho, Otsu, Shiga, 520-2192, Japan}

\author{Shogo Otsubo}
\affiliation{Laboratory of Infrared High-resolution spectroscopy(LiH), Koyama Astronomical Observatory, Kyoto Sangyo University, Motoyama, Kamigamo, Kita-ku, Kyoto 603--8555, Japan}

\author[0000-0002-5756-067X]{Akira Arai}
\affiliation{Subaru Telescope, National Astronomical Observatory of Japan, 650 North A’ohoku Place, Hilo, HI 96720, USA}
\affiliation{Laboratory of Infrared High-resolution spectroscopy(LiH), Koyama Astronomical Observatory, Kyoto Sangyo University, Motoyama, Kamigamo, Kita-ku, Kyoto 603--8555, Japan}

\author[0000-0003-3579-7454]{Chikako Yasui}
\affiliation{National Astronomical Observatory of Japan, 2-21-1 Osawa, Mitaka, Tokyo 181--8588, Japan}
\affiliation{Laboratory of Infrared High-resolution Spectroscopy (LiH), Koyama Astronomical Observatory, Kyoto Sangyo University, Motoyama, Kamigamo, Kita-ku, Kyoto 603--8555, Japan}

\author{Hitomi Kobayashi}
\affiliation{Kyoto Nijikoubou, LLP, 17-203, Iwakura-Minamiosagi-cho, Sakyo-ku, Kyoto 606--0003, Japan}

\author[0000-0002-4896-8841]{Giuseppe Bono}
\affiliation{Dipartimento di Fisica, Universit\`{a} di Roma Tor Vergata, via della Ricerca Scientifica 1, I-00133 Roma, Italy}
\affiliation{INAF-Osservatorio Astronomico di Roma, via Frascati 33, I-00078 Monte Porzio Catone, Italy}

\author[0000-0002-5878-5299]{Ivo Saviane}
\affiliation{European Southern Observatory, Alonso de Cordova 3107, Santiago, Chile}

\begin{abstract}
We searched for diffuse interstellar bands (DIBs) in the 0.91$<\lambda<$1.33 $\mu$m region by analyzing the near-infrared (NIR) high-resolution ($R=20,000$ and 28,000) spectra of 31 reddened early-type stars ($0.04<E(B-V)<4.58$) and an unreddened reference star. The spectra were collected using the WINERED spectrograph, which was mounted on the 1.3 m Araki telescope at Koyama Astronomical Observatory, Japan, in 2012--2016, and on the 3.58 m New Technology Telescope at La Silla Observatory, Chile, in 2017--2018. We detected \textcolor{black}{54} DIBs --- \textcolor{black}{25} of which are newly detected by this study --- \textcolor{black}{eight} DIB candidates. 
Using this updated list, the DIB distributions over a wide wavelength range from optical to NIR are investigated. 
The FWHM values of the NIR DIBs are found to be narrower than those of the optical DIBs on average, which suggests that the DIBs at longer wavelengths tend to be caused by larger molecules. 
Assuming that the larger carriers are responsible for the DIBs
at longer wavelengths and have the larger oscillator strengths,
we found that the total column densities of the DIB carriers tend to
decrease with increasing DIB wavelength. 
The candidate molecules and ions for the NIR DIBs are also discussed. 

\end{abstract}


\section{Introduction}

Diffuse interstellar bands (DIBs) are ubiquitous absorption features that are detected in the visible to near-infrared (NIR) spectra of reddened stars. Their absorbing matter is considered to be gas-phase carbonaceous molecules such as carbon chain molecules, polycyclic aromatic hydrocarbons (PAHs), and fullerenes. However, there is no critical evidence to confirm this. 
\textcolor{black}{The ionized buckminsterfullerene (C$_{60}^+$) was almost confirmed as the carrier of the five NIR DIBs via a comparison between the astronomical spectra and the spectrum of the gas-phase C$_{60}^+$ obtained from laboratory experiments \citep{cam15,wal16,cam18,cor19}, although the identification is still under discussion, because the intensity ratios of the two strong bands were quite variable after the correction of the contaminated stellar line and the detection of the weak bands was not robust \citep{gal17a,gal21}.}
The gas-phase absorption spectra of some neutral PAH molecules have also been measured and compared to DIBs \citep{gre11,sal11}. Although this comparison did not result in the detection of corresponding absorption bands in the astronomical spectra, \citet{gre11} and \citet{sal11} succeeded in setting stringent upper limits for the abundances of specific neutral PAHs in some translucent clouds. Further laboratory experiments are required to test the so-called PAH--DIB hypothesis \citep{leg85,van85}.

Spectroscopic surveys of DIBs with high signal-to-noise (S/N) ratios over a wide range of wavelengths are essential for the identification of DIB carriers via a comparison of the DIB spectra with experimentally obtained absorption spectra of the candidate molecules. Some spectroscopic surveys in the optical wavelength range have successfully detected hundreds of DIBs within a detection limit of a few m\r{A} in equivalent widths (EWs) \citep{jen94,wes00,tua00,gal00,hob08,hob09,fan19}. However, DIB observations in the NIR range have been limited, because the performance of NIR high-resolution spectrographs (in terms of spectral resolution and sensitivity) has been lower than that of optical spectrographs. In addition, many telluric absorption bands in the NIR range prevent  the detection of weak absorption features. Recent progress in NIR spectroscopy has enabled searches for DIBs in the NIR spectra over the last decade \citep{geb11,cox14,ham15,ely17,gal17b,lal18}.

We have conducted a comprehensive survey of NIR DIBs using the high-sensitivity high-resolution NIR WINERED spectrograph \citep{ike16}. In our first study \citep[][hereafter, H15]{ham15}, we observed 25 stars and successfully identified 15 new DIBs in the range of $0.91<\lambda<1.32$ $\mu$m. Moreover, we found that the EWs of some NIR DIBs are highly correlated with each other, which suggests that their carriers have similar molecular properties. In our subsequent study \citep[][hereafter, H16]{ham16}, we investigated the environmental dependence of NIR DIB strengths using the high-quality spectra of seven bright stars in the Cygnus OB2 association, which is one of the most massive clusters or OB associations veiled with large interstellar extinction \citep[the extinction of the most reddened member, No.\,12, reaches $A_V =10.2$ mag; ][]{wri15,whi15}. Owing to the large extinction and high flux density in the NIR wavelength range, we could detect even the weakest DIBs reported in H15 with high precision. We found that the NIR DIBs are not correlated with the column densities of the line-of-sight C$_2$ molecules, which suggests that the DIB carriers are not distributed in the molecular clouds traced with C$_2$ molecules. We also found that the carrier of DIB $\lambda 10504$ would be destroyed by the strong UV radiation in the Cyg OB2 association, whereas the DIB carriers of other strong DIBs survive in this environment, which suggests differences in carrier properties, such as the ionization potential and dissociation energy.

In this series, we explore the properties of the DIBs in the $Y$ and $J$ bands, using the large quantity of high-quality WINERED data. The most important DIBs, the C$_{60}^+$ bands, which are covered by the spectrograph, will be comprehensively investigated in order to reveal the properties of C$_{60}^+$ in the interstellar medium. It is also of great interest to explore the relation of other DIBs with C$_{60}^+$. We will also study the correlations between NIR DIBs and optical DIBs, which have been extensively investigated. Through this series, we aim to constrain the carriers of DIBs in the $Y$ and $J$ bands. 

In this first study of the series, we updated the catalog of DIBs in the 0.91--1.33 $\mu$m range, using a larger sample and higher quality of spectra than those of H15. We analyzed the high-S/N spectra of 32 objects, comprising of an unreddened reference star ($\beta$ Ori) and 31 reddened stars in the range $0.04 < E(B-V) < 4.58$. To find very weak DIBs, we included large-extinction lines of sight of the Cyg OB2 association \citep[$2.2<A_V<10.2$ mag;][]{wri15} and Westerlund 1 (Wd 1) cluster \citep[$8.5<A_V<17$ mag;][]{dam16}. In comparison with H15, the sensitivity and wavelength stability of WINERED were improved by the upgrade \citep{ike22}. The accuracy of the removal of telluric absorption lines is also improved \citep{sam18}. With these improvements, we can detect much weaker DIBs than those in H15. The updated catalog of DIBs in the $Y$ and $J$ bands will be the basis of this series of papers.

The remainder of this paper is organized as follows. In Section 2, we describe our observations and targets. In Section 3, our data reduction procedures are described. In Section 4, we describe the newly detected DIBs, and in Section 5, we discuss the distributions of DIBs from the optical wavelength to the NIR range and the carriers of the newly found DIBs. Finally, we present a summary of the study in Section 6.



\section{Observations and Targets}
\subsection{Observations}


The data were collected using the high-resolution NIR echelle spectrograph, WINERED \citep{ike16,ike22}, which uses a 1.7 $\mu$m cutoff 2048 $\times$ 2048 HAWAII-2RG IR array. We used the WIDE mode, with a 100 $\mu$m slit, which includes the wavelength range of 0.91--1.35 $\mu$m, with a spectral resolving power of $R\equiv \lambda/\Delta \lambda =$ 28,000 or $\Delta v =11$ km s$^{-1}$.

WINERED was mounted on the F/10 Nasmyth focus of the 1.3-m Araki telescope at Koyama Astronomical Observatory, Kyoto Sangyo University, Japan \citep{yos12}, from 2012 to 2016 December. From 2017 to 2018, WINERED was mounted on the ESO 3.58 m New Technology Telescope (NTT) at the La Silla Observatory, Chile. The pixel scales of WINERED were 0$''$.8 pixel$^{-1}$ and 0$''$.27 pixel$^{-1}$ for the Araki telescope and the NTT, respectively. Most of the data were obtained with the telescope dithered by 30 arcsec (Araki telescope) or 10 arcsec (NTT) --- the so-called ``ABBA" sequence. In the few cases in which the seeing was not good, we alternately obtained the spectra of the object and sky (the ``OSO" sequence). The telluric-standard A0V--A2V type stars were observed at airmasses that were similar to those of the targets. 

The observations made with the Araki telescope were primarily conducted in 2014 January and 2014 August--October. Only the observation of $\beta$ Ori was conducted on 2016 February 11. The typical seeing size at the Koyama Astronomical Observatory is approximately 3--5 arcsec. Some of the data have already been published in H15 and H16. Immediately prior to conducting the observations in 2014 August, we installed an $H$-band blocking filter to decrease the thermal leak in the 1.7--1.8 $\mu$m range. The new filter successfully decreased the background noise, and it yielded spectra with high-S/N ratios. However, the filter was bent by the tight mechanical mounting, which caused a slight off-focus of the light on the IR array. Consequently, the spectral resolving power was reduced to $R=20,000$, which is lower than the nominal value, for the observations made during 2014 August--October (since then, the problem has been solved). 

\begin{figure}
\includegraphics[width=8cm,clip]{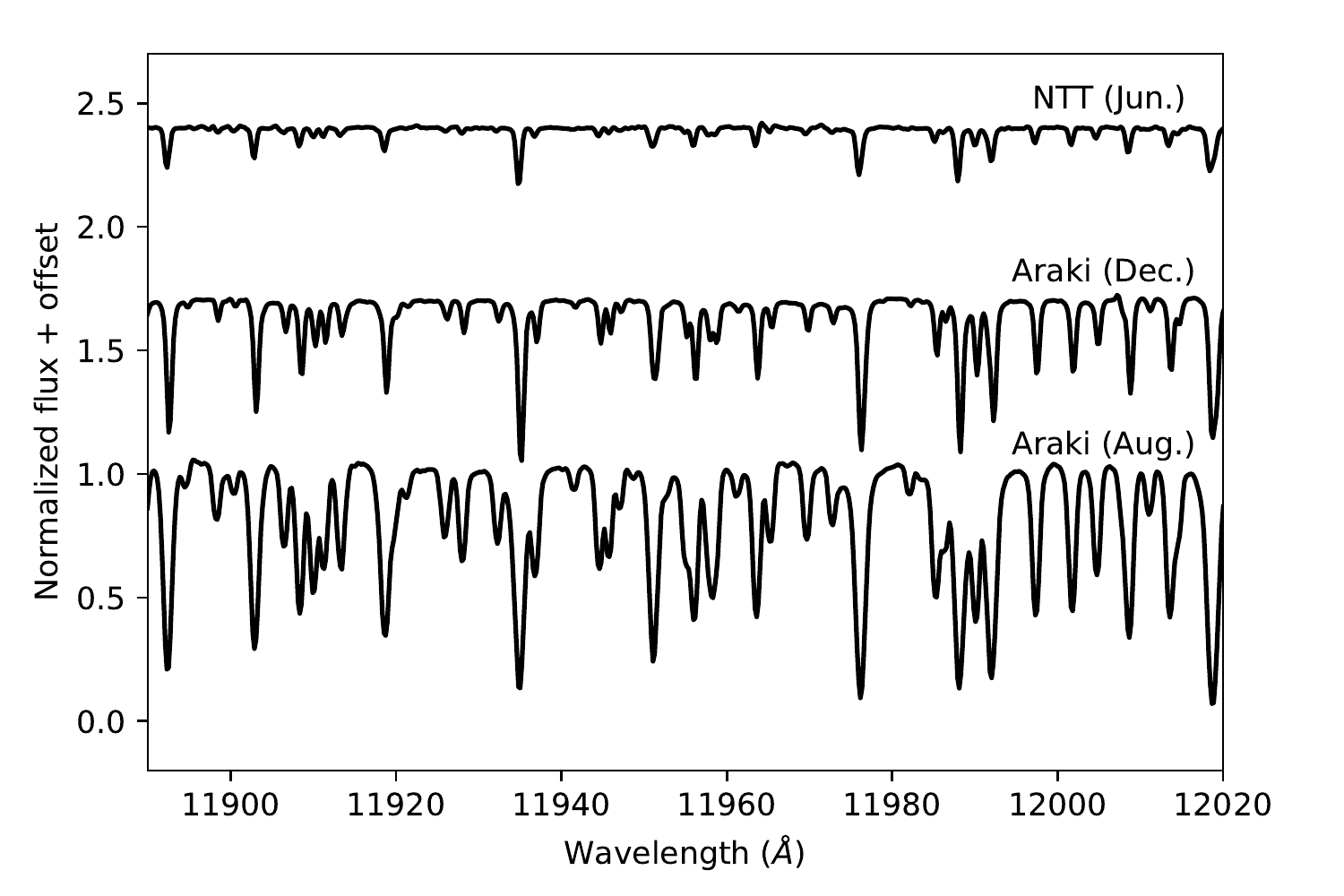}
\caption{Variations of the telluric lines with the observation sites and seasons. The normalized WINERED spectra of the telluric-standard stars obtained using the NTT in June (top), the Araki telescope in December (middle), and the Araki telescope in August (bottom) are shown.}
\label{telsite}
\end{figure}

The observations using the NTT were conducted in 2017 July (ESO program ID: 099.C-0850(B)). The typical seeing size was approximately 1 arcsec. Compared with the spectra obtained using the Araki telescope, the telluric absorption lines of the NTT spectra were considerably weaker, owing to the lower humidity and higher elevation of the La Silla Observatory. Figure \ref{telsite} shows the comparison of the spectra of the telluric-standard stars. For the spectra obtained using the Araki telescope, the telluric absorption lines strongly depend on the season in which the observations were made. Very strong absorption lines of atmospheric water vapor can be seen in the Araki spectra obtained in August, which is the season of high humidity and temperatures in Japan. The difference in the intensities of the telluric absorption lines affected on the DIB search. The wavelength ranges in which the water vapor lines are strong, such as $0.93<\lambda<0.95$ $\mu$m, $1.11<\lambda<1.17$ $\mu$m and $1.33<\lambda<1.35$ $\mu$m, were not available for the DIB search in the Araki spectra obtained in August.

\subsection{Targets}

\begin{figure}
\includegraphics[width=8cm,clip]{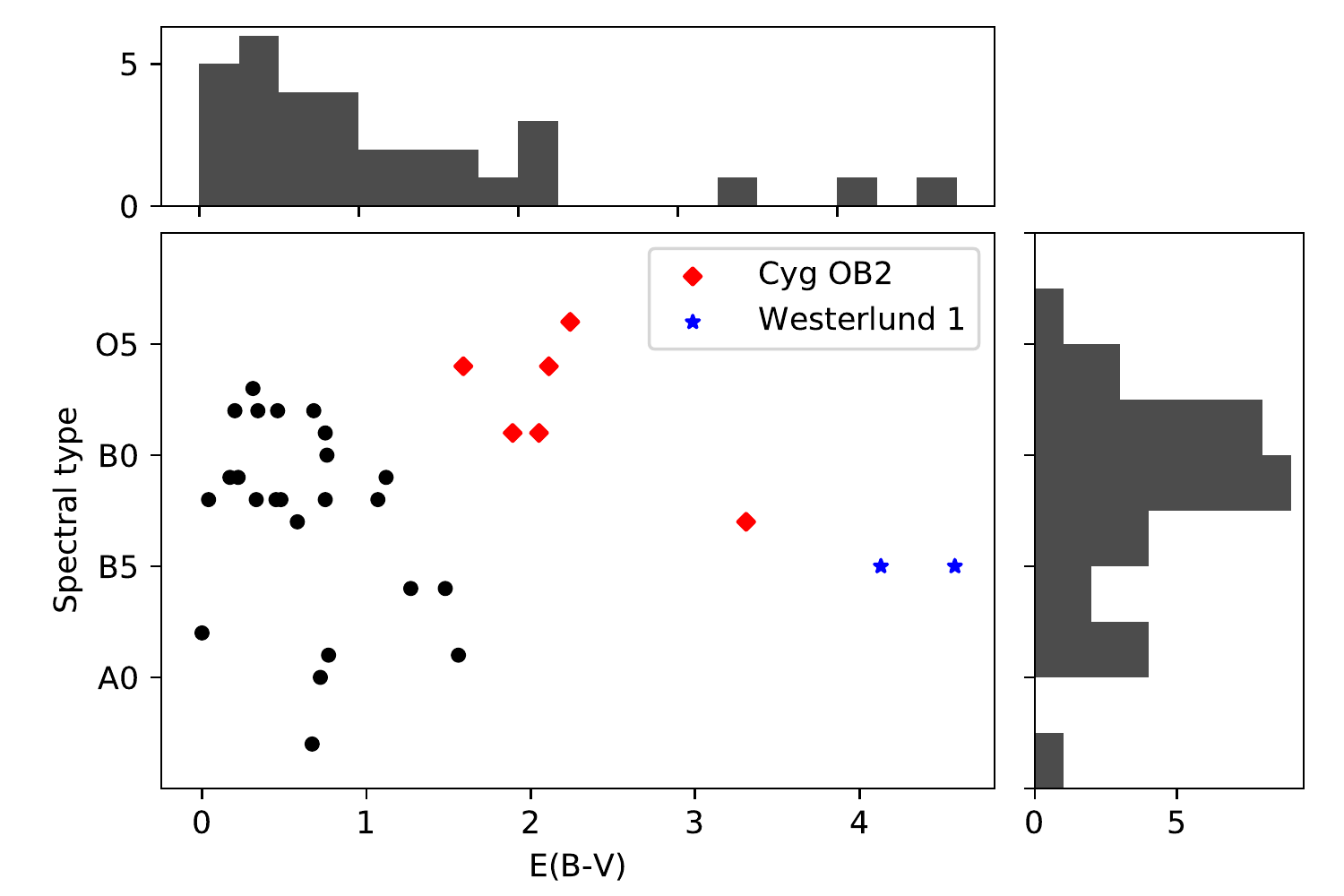}
\caption{$E(B-V)$ distribution and the spectral types of our targets.}
\label{ebvsptype}
\end{figure}

The targets and observation information are shown in Table \ref{targets}. We observed 31 reddened early-type stars and one 1 reference star without any interstellar extinction ($\beta$ Ori). Figure \ref{ebvsptype} shows the distribution of $E(B-V)$ and the spectral types of our targets. The reddened stars cover a wide range of spectral types, from A3 to O4.5; however, our observations lack of the spectral-type coverage of the reference star: only $\beta$ Ori (B8Iab). To avoid the misidentification of the stellar absorption lines as the DIBs, we also utilized the model spectrum synthesized using SPTOOL (Y. Takeda, private communication) as a reference for the stellar spectra. Our targets include a wide range of color excesses, $0.04 < E(B-V) < 4.58$, which enable us to find the DIBs that originate from interstellar clouds. We comment on some targets below.

HD183143 is the most extensively observed star for DIB studies, owing to its large flux density and DIB strength, which is a result of its large extinction ($E(B$ - $V)=1.27$). The DIB properties of this star are well known, owing to previous surveys of DIBs in optical \citep[e.g.,][]{her75,her91,jen94,tua00,gal00,hob09,cox14} and NIR ranges \citep{cox14}. \textcolor{black}{The distributions and properties of DIBs over a wide wavelength range are discussed in Section 5.1 using the data of HD183143.} 

The Cyg OB2 association is one of the most massive clusters in our galaxy. We examined the spectra of six stars from Cyg OB2, which were also investigated in H16. The distance to the Cyg OB2 is estimated as 1669 $\pm$ 6 pc based on Gaia Data Release 2 data \citep{ore21}, and it has a very large extinction \citep{wri15}. In H16, we detected DIBs with very large EWs in the spectra of the Cyg OB2 members. In addition, various gaseous environments can be traced using this target, because the gas clouds in the lines of sight of Cyg OB2 have a complex structure, consisting of a diffuse component and a patchy dense component \citep{whi15}. Therefore, Cyg OB2 is an ideal target for finding very weak DIBs.

Wd 1 is a massive young stellar cluster with a mass of $\sim 10^5 M_\odot$ \citep{neg10} located at a distance of $d = 4.12^{+0.66}_{-0.33}$ kpc from the Sun \citep{bea21}. The typical extinction of the cluster members is  $A_V = 11.4 \pm 1.2$ \citep{dam16}. W7 and W33, which are our targets, are very luminous B supergiants in the cluster \citep{cla05}. The color excesses of these two stars were calculated by using the stellar intrinsic colors of \citet{cox00}. Owing to these color excesses being the highest among our targets, the DIBs are the strongest; thus, these stars of Wd 1 are the best targets for detecting new weak DIBs.

\movetabledown=5cm
\begin{rotatetable}
\begin{deluxetable*}{ccccccccccccc}
\tabletypesize{\scriptsize}
\tablecaption{Summary of Targets \label{targets}}
\tablehead{
 \colhead{Object} & \colhead{} & \colhead{} & \colhead{} & \colhead{}  &\colhead{Telluric \tablenotemark{a}} &\colhead{} &\colhead{} &\colhead{} &\colhead{S/N\tablenotemark{b}}  & \colhead{Obs. Date} & \colhead{Telescope} & \colhead{$R$} \\
 \colhead{Name} & \colhead{Sp. Type} & \colhead{$J$ (mag)} & \colhead{$E(B-V)$} & \colhead{Int. Time (s)} & \colhead{Name} & \colhead{Sp. Type} & \colhead{$J$ (mag)} & \colhead{Int. Time (s)} & \colhead{} & \colhead{(UT)} & \colhead{} & \colhead{} 
}
\startdata
$\beta$ Ori & B8Iae & 0.220 & 0.00 & 150 & HR 1683 & A0V & 5.829 & 6000 & 700 & 2016/02/11 & Araki & 28000 \\
HD 210191 & B2III & 6.128 & 0.04 & 900 & HR 8134 & A1V & 6.284 & 810 & 350 & 2017/07/30 & NTT & 28000 \\
HD 141637 & B1.5Vn & 4.802 & 0.17 & 240 & HD 139129 & A0V & 5.392 & 480 & 800 & 2017/07/29 & NTT & 28000 \\
HD 135591 & O8IV((f)) & 5.580 & 0.20 & 360 & HD 139129 & A0V & 5.392 & 480 & 600 & 2017/07/29 & NTT & 28000 \\
HD 144470 & B1V & 4.000 & 0.22 & 120 & HD 139129 & A0V & 5.392 & 480 & 500 & 2017/07/29 & NTT & 28000 \\
HD 155806 & O7.5V((f))z(e) & 5.655 & 0.31 & 240 & HD 163336 & A0V & 5.8 & 480 & 400 & 2017/07/30 & NTT & 28000 \\
HD 179406 & B2/3II & 5.060 & 0.33 & 120 & HR 8134 & A1V & 6.284 & 810 & 400 & 2017/07/30 & NTT & 28000 \\
HD 151804 & O8Iaf & 4.939 & 0.34 & 120 & HR 7159 & A1V & 5.962 & 400 & 500 & 2017/07/29 & NTT & 28000 \\
HD 41117 & B2Ia & 4.040 & 0.45 & 720 & 21 Lyn & A0.5V & 4.695 & 1200 & 400 & 2014/01/23 & Araki & 28000 \\
HD 152408 & O8Iape & 5.210 & 0.46 & 360 & HD 139129 & A0V & 5.392 & 600 & 400 & 2017/07/30 & NTT & 28000 \\
HD 170740 & B2/3II & 5.222 & 0.48 & 120 & HD 163336 & A0V & 5.8 & 480 & 350 & 2017/07/30 & NTT & 28000 \\
HD 43384 & B3Iab & 5.187 & 0.58 & 1440 & 21 Lyn & A0.5V & 4.695 & 1200 & 400 & 2014/01/23 & Araki & 28000 \\
HD 223385 & A3Iae & 3.866 & 0.67 & 720 & 50 Cas & A2V & 3.885 & 720 & 550 & 2014/01/22 & Araki & 28000 \\
HD 149404 & O8.5Iab(f)p & 4.606 & 0.68 & 120 & HD 163336 & A0V & 5.8 & 720 & 350 & 2017/07/29 & NTT & 28000 \\
HD 20041 & A0Ia & 4.057 & 0.72 & 1200 & 50 Cas & A2V & 3.885 & 1800 & 750 & 2014/01/21 & Araki & 28000 \\
HD 154368 & O9.5Iab & 5.024 & 0.75 & 240 & HD 139129 & A0V & 5.392 & 600 & 550 & 2017/07/30 & NTT & 28000 \\
HD 148379 & B2Iab & 4.050 & 0.75 & 240 & HD 139129 & A0V & 5.392 & 600 & 700 & 2017/07/30 & NTT & 28000 \\
HD 152235 & B0.5Ia & 5.030 & 0.76 & 120 & HR 7159 & A1V & 5.962 & 400 & 400 & 2017/07/29 & NTT & 28000 \\
HD 185247 & B9V & 7.110 & 0.77 & 600 & HR 7159 & A1V & 5.962 & 400 & 400 & 2017/07/29 & NTT & 28000 \\
HD 147889 & B2III/IV & 5.344 & 1.07 & 480 & HD 163336 & A0V & 5.8 & 720 & 550 & 2017/07/29 & NTT & 28000 \\
HD 169454 & B1Ia & 4.450 & 1.12 & 120 & HD 163336 & A0V & 5.8 & 480 & 400 & 2017/07/30 & NTT & 28000 \\
HD 183143 & B6Ia & 4.130 & 1.27 & 480 & 35 Vul & A1V & 5.276 & 1800 & 500 & 2014/08/22 & Araki & 20000 \\
  &  &  &  & 600 & c And & A0V & 5.254 & 1200 & 400 & 2014/09/20 & Araki & 20000 \\
HD 168625 & B6Iap & 5.070 & 1.48 & 180 & HD 163336 & A0V & 5.8 & 480 & 450 & 2017/07/30 & NTT & 28000 \\
HD 168607 & B9Iaep & 4.510 & 1.56 & 180 & HD 163336 & A0V & 5.8 & 480 & 500 & 2017/07/30 & NTT & 28000 \\
Cyg OB2 & & & & & & & & & & & & \\
No.\,8A & O6Ib(fc)+O4.5:III:(fc) & 6.123 & 1.59 & 2400 & 35 Vul & A1V & 5.276 & 1200 & 330 & 2014/09/16 & Araki & 20000 \\
No.\,10 & O9.7Iab & 6.294 & 1.89 & 1800 & $\rho$ Peg & A1V & 4.9 & 1200 & 250 & 2014/09/11 & Araki & 20000 \\
No.\,3 & O9: & 6.498 & 2.05 & 1800 & $\rho$ Peg & A1V & 4.9 & 1200 & 360 & 2014/09/11 & Araki & 20000 \\
No.\,5 & O6.5-7f+O5.5-6f & 5.187 & 2.11 & 1200 & 29 Vul & A0V & 5.153 & 1200 & 360 & 2014/08/30 & Araki & 20000 \\
  &  &  &  & 1200 & 35 Vul & A1V& 5.276 & 900 & 150 & 2014/09/09 & Araki & 20000 \\
  &  &  &  & 1200 & c And & A0V & 5.254 & 1200 & 380 & 2014/09/20 & Araki & 20000 \\
No.\,9 & O4.5If & 6.468 & 2.24 & 1800 & 29 Vul & A0V & 5.153 & 1200 &300 & 2014/09/13 & Araki & 20000 \\
No.\,12 & B3-4Ia+ & 4.667 & 3.31 & 3600 & HR 196 & A2V & 5.25 & 8400 & 600 & 2014/10/17 & Araki & 20000 \\
Westerlund 1 & & & & & & & & & & & & \\
W 7 & B5Ia+ & 6.847 & 4.13 & 1800 & HD 139129 & A0V & 5.392 & 600 & 200 & 2017/07/30 & NTT & 28000 \\
W 33 & B5Ia+ & 6.883 & 4.58 & 1500 & HD 163336 & A0V & 5.8 & 720 & 400 & 2017/07/29 & NTT & 28000 \\
\enddata
\tablecomments{}
\tablenotetext{a}{Telluric-standard stars used for the correction of the telluric absorption lines.}
\tablenotetext{b}{The S/N ratio per pixel after the division by a telluric-standard spectrum at the center of the $Y$ band.}
\end{deluxetable*}
\end{rotatetable}

\newpage 
\ \\ \ 
\newpage

\section{Analysis}
\subsection{Data reduction}


The obtained raw data were processed using the pipeline that was developed to reduce the WINERED data. The pipeline automatically produces the wavelength-calibrated one-dimensional spectra from the raw data. The pipeline conducts the subtraction of sky frames, the subtraction of scattered light, flat-fielding, bad pixel interpolation, the correction of echellogram distortion, spectrum extraction, wavelength calibration, and continuum normalization. A brief description of the pipeline software is provided in \citet{ike22}. A more detailed description will be given  by S. Hamano et al., in preparation.

Since most of the optical components of the WINERED spectrograph are maintained at room temperature, the wavelength of the spectrum of each frame shifted slightly with the changes in ambient temperature during the observations. The relative wavelength shifts among the multiple frames of a target are corrected in the pipeline by aligning the wavelengths of the input frames with that of the frame with the highest count. However, the absolute wavelength shift from the dispersion solution was left in the pipeline-reduced spectra. In this study, the shift was measured using the cross-correlation between the model spectra of telluric absorption and the pipeline-reduced spectra. Because we obtained high-S/N spectra of early-type stars, for which the stellar lines are fewer than for late-type stars, we could measure the wavelength shifts with a high accuracy. The wavelengths of the spectra were recalibrated with the shift values.

The telluric lines of the target spectrum were removed using the method detailed in \citet{sam18}. In the method, we made synthetic telluric spectra using \textbf{molecfit} \citep{sme15,kau15}. Using these spectra as a reference, we identified stellar features in the standard spectrum and fitted them with multiple Gaussian curves, which were synthesized to construct a stellar absorption-line spectrum. \textcolor{black}{Because the telluric-standard stars are A0V--A2V dwarf stars, the metal lines of telluric-standard stars are not complex and can be fitted well with multiple Gaussian curves.} The observed standard spectra were divided by the synthesized spectrum, to cancel out intrinsic stellar features, and the resulting spectra were used to correct the telluric absorption lines of the targets. A detailed description of this procedure is provided in \citet{sam18}.

\subsection{DIB measurement}

%
Before the search for new DIBs, we measured the parameters of the DIBs that were found using the originally developed software. The spectrum files, the rest-frame DIB wavelengths, and the guess velocities of the DIBs were inputted to the software. First, the software cut out the input spectrum around the DIBs in the range of $\pm 500$ km s$^{-1}$ and normalized it by fitting a Legendre function to the surrounding continuum region. Then, the S/N ratio per pixel was calculated from the standard deviation of the continuum region. Using the S/N ratio and the normalized spectrum of the telluric-standard star, the uncertainty of each pixel of the input spectrum was calculated. The absorption regions, for which the depths by the peak are lower than the continuum level (=1) by 2$\sigma$, are then automatically searched for in the normalized spectrum. The region in which the center velocity is the closest to the input velocity is picked up as the DIB region. If the absorption region cannot be found within a $\pm$50 km s$^{-1}$ range from the input velocity, the DIB is judged as a nondetection. 

For the detected DIBs, the EWs and their uncertainties, the central wavelengths and corresponding heliocentric velocities, the FWHM in a wavelength scale, the wavelengths of the absorption peak, and the depth were measured. For the undetected DIBs, the 3$\sigma$ upper limits of the EWs are calculated. The DIB EWs ($W$) were measured from the normalized spectra as follows:
\begin{equation}
W = \sum_{i=1}^N \left( 1- I_n(x_i) \right) \Delta \lambda  ,
\end{equation}
where $x_i$ is the $i$ th pixel, $N$ is the total number of summed pixels, $I_n(x_i)$ is the normalized flux at $x_i$, and $\Delta \lambda$ is the wavelength width per pixel. The EW uncertainty was then calculated using the following equation:
\begin{equation}
\sigma_W = (\sigma_\text{stat}^2 + \sigma_\text{cont}^2)^{1/2}, \label{ewerr}
\end{equation}
where $\sigma_\text{stat}$ and $\sigma_\text{cont}$ are the statistical uncertainty and the systematic uncertainty from continuum fitting, respectively. $\sigma_\text{stat}$ was calculated as follows:
\begin{equation}
\sigma_\text{stat} = \left( \sum_{i=1}^N  \delta I_n^2(x_i) \right)^{1/2} \Delta \lambda, 
\end{equation}
where $\delta I_n(x_i)$ is the uncertainty of the normalized flux at $x_i$. $\sigma_\text{cont}$ was estimated with the rms shift method, using the statistical uncertainties \citep{sem92}. In the case of a nondetection, the upper limit of the EW is calculated using $3 \sigma_W$ with Eq. (\ref{ewerr}), by setting the integration range from the typical width of the DIBs and the input velocity. 

The central wavelength of each band, $\lambda _c$, was measured as the weighted average for the overall band as follows:
\begin{equation}
\lambda _c = \frac{\sum_{i=1}^N \lambda _i \tau (x_i)}{\sum_{i=1}^N \tau (x_i)}, \label{lam_c}
\end{equation}
where $\tau (x_i)$ is the optical depth at $x_i$.
A formulation was adopted from a preceding survey of DIBs in the optical wavelength range in \citet{fan19}. For a DIB with an asymmetric profile, the central wavelength measured using Eq. \ref{lam_c} differs from the wavelengths at the absorption peak. 
The FWHMs of each DIB were determined by calculating the difference between the wavelengths where the depths are half of the peak depth. 

All the results of the continuum normalization, the search for DIBs, and the DIB measurements were checked by eye, and if some of the procedures had failed, owing to the systematic noise by telluric absorption lines and the blending of the other features, the fitting parameters of the continuum normalization and/or the integration range were adjusted appropriately.

\begin{figure*}
\includegraphics[width=16cm,clip]{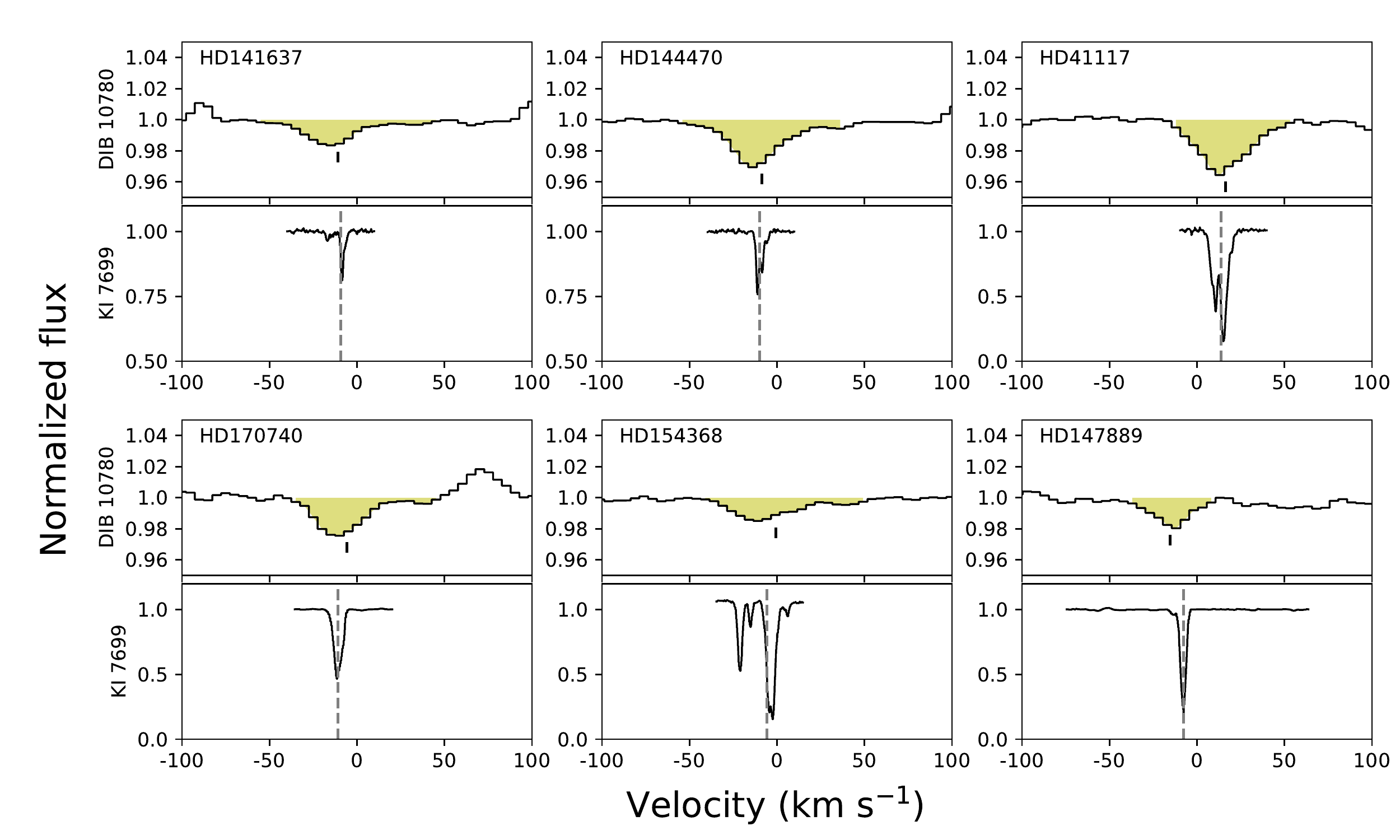}
\caption{\textcolor{black}{Comparison of the spectra of DIB $\lambda$10780 (this study) and \ion{K}{1} 7699\r{A} for six targets, which were used to determine the rest-frame wavelength of DIB $\lambda$10780. The spectra of the \ion{K}{1} absorption lines were reproduced from \citet{wel01} (for HD41117, HD141637, HD144470, and HD154368) and \citet{sie20} (for HD170740 and HD147889). The integration ranges of the DIB absorption are shown by the yellow areas. The central velocities measured from the profile of DIB 10780 are marked with the  vertical black lines. The \ion{K}{1} velocities weighted with the column densities of each velocity component are shown by the gray dashed lines.}}
\label{KIprofile}
\end{figure*}

The velocities of the line-of-sight interstellar clouds are necessary for the search for the DIBs in the spectrum. We do not have reliable interstellar features in the $0.91<\lambda<1.33$ $\mu$m range to determine the line-of-sight velocities of the clouds, such as the \ion{K}{1} line at $\lambda=7698.9645$ \r{A}, which has frequently been used to investigate the velocity profiles of the interstellar absorption. In this study, we use the DIB $\lambda$10780, which is relatively narrow and strong in this wavelength range; therefore, it is likely to be best suited for the velocity measurement. First, the rest-frame wavelength of DIB $\lambda 10780$ is determined by comparing the velocities calculated by DIB $\lambda$10780 with those measured using the \ion{K}{1} line. Among our 31 targets, the \ion{K}{1} line profiles of 10 targets have been obtained in previous studies. We used the average of the velocities of the components, weighted by the column densities for the targets toward which multiple-velocity components are detected using a \ion{K}{1} line. 
Table \ref{velocity} shows the velocities and references of the \ion{K}{1} velocities. We determined the rest-frame wavelength of DIB $\lambda$10780 as $\lambda_\text{rest}=10780.6$\r{A} by minimizing the difference between the \ion{K}{1} velocities and the $\lambda 10780$ velocities. 
In the calculation, we used only six targets, for which the FWHMs of DIB $\lambda$10780 are lower than 1.5\r{A}, in order to reduce the systematic uncertainty in the rest-frame wavelength, owing to the blending of multiple-velocity components. \textcolor{black}{The FWHMs of DIB $\lambda$10780 are also listed in Table \ref{velocity}. Figure \ref{KIprofile} shows the comparison between the spectra of DIB $\lambda$10780 and \ion{K}{1} 7699 for the six targets.}

Our value, $10780.6$\r{A}, is slightly larger than the previous values of $10780.3\pm0.2$\r{A} \citep{cox14} and $10780.46\pm0.10$\r{A} (H15), which were determined using a Gaussian fitting. The difference in the rest-frame wavelength is caused by the difference in the method of the DIB wavelength measurements because the $\lambda10780$ profile is asymmetric. We compared the wavelengths measured using our method (Eq. \ref{lam_c}) with a Gaussian fitting. It was found that the central wavelengths of DIB $\lambda10780$ measured using a Gaussian fitting are smaller than those calculated with Eq. (\ref{lam_c}) by $\sim 0.2$\r{A}, which is comparable to the difference between the rest-frame wavelength of $\lambda$10780 in this study and those in the previous studies.

\begin{figure*}
\includegraphics[width=16cm,clip]{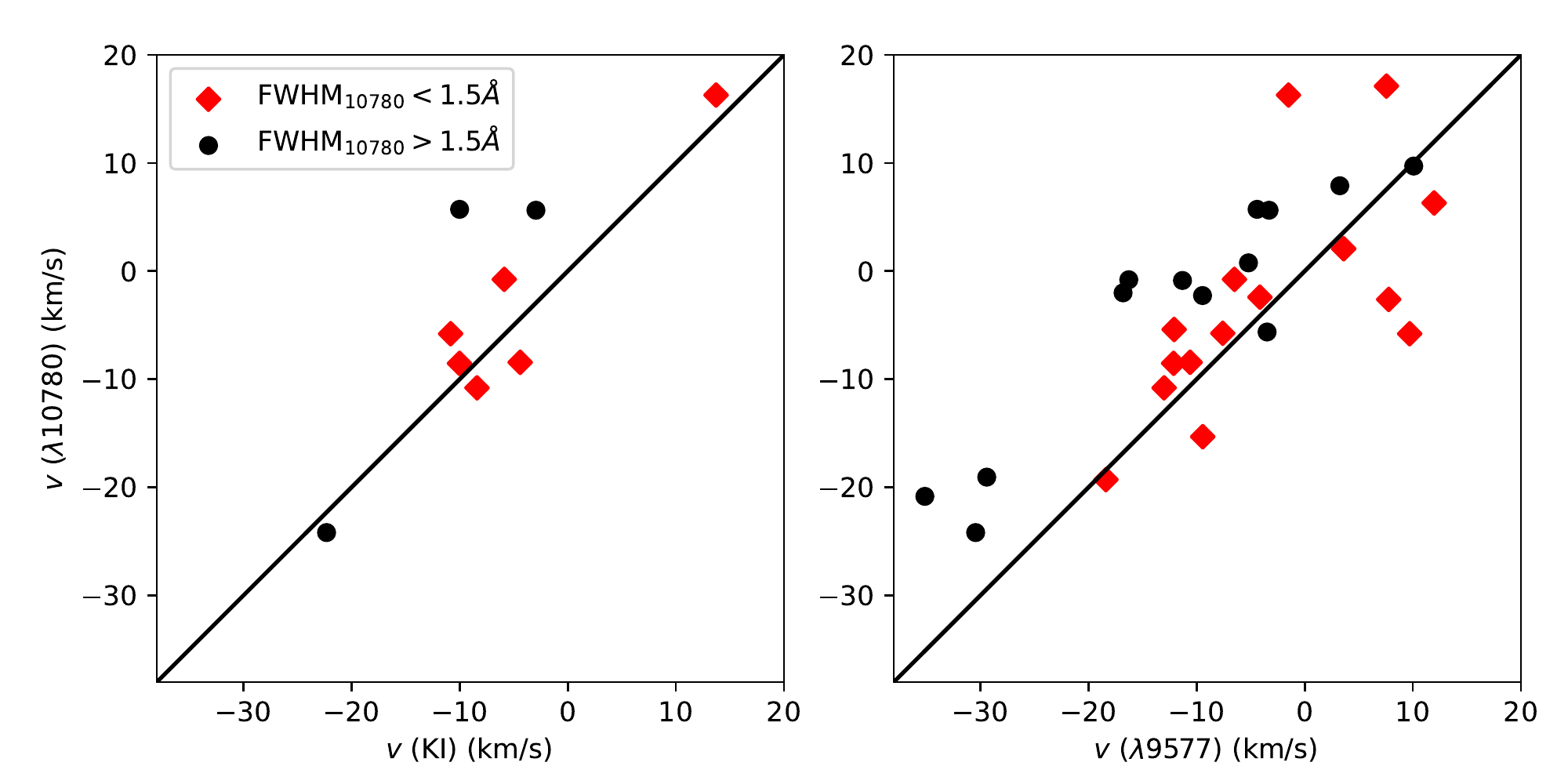}
\caption{Comparison between the velocities determined by the DIB $\lambda$10780 ($\lambda_\text{rest}=10780.6$\r{A}) and those with a \ion{K}{1} line (left) and the C$_{60}^+$ DIB $\lambda 9577$ (right). The stars with a FWHM$_\text{10780} < 1.5$ \r{A} are shown with red diamonds, and those with a FWHM$_\text{10780} < 1.5$ \r{A} are shown with black circles.}
\label{velocitycomp}
\end{figure*}

We further checked the robustness of the rest-frame wavelength of DIB $\lambda 10780$ by comparing the velocities measured using a C$_{60}^+$ band, DIB $\lambda 9577$. Although DIB $\lambda 9577$ is not suited for a velocity measurement, owing to the contamination of the strong telluric absorption lines and the intrinsically broad profile, the rest-frame wavelength is measured in the laboratory experiments as $9577.0\pm0.2$\r{A} \citep{cam18}. \textcolor{black}{\citet{gal17a} measured the rest-frame wavelengths of C$_{60}^+$ bands $\lambda\lambda$9577 and 9632 for 19 reddened stars, and showed that the mean wavelengths, 9577.0 and 9632.2 \r{A}, matched with the wavelengths measured in the laboratory experiments but with a large scatter.} The right panel of Figure \ref{velocitycomp} shows the comparison between the velocities. It was found that the velocities measured using $\lambda10780$ and $\lambda9577$ are consistent, which suggests that the rest-frame wavelength of $\lambda$10780, $10780.6$\r{A}, is robust for determining the line-of-sight velocities of the interstellar clouds of our targets. 


\begin{deluxetable}{ccccc}
\tabletypesize{\scriptsize}
\tablecaption{Velocities of the interstellar clouds}
\tablehead{
 \colhead{Object}  & \colhead{$v_\text{\ion{K}{1}}$} & \colhead{$v_{\lambda 10780}$} & \colhead{FWHM$_{\lambda 10780}$ } & \colhead{Reference\tablenotemark{a}}  \\
 \colhead{} &  \colhead{km s$^{-1}$} &  \colhead{km s$^{-1}$} &  \colhead{\r{A}} &  \colhead{} 
}
\startdata
HD 210191   &  & ... & ... & \\
HD 20041  & & -5.4 & 1.3 & \\
HD 141637\tablenotemark{c} & -9.40 & -10.8 & 1.1 & 1 \\
HD 135591  & & 6.3  & 1.2 & \\
HD 144470\tablenotemark{c}  & -10.01 & -8.5 & 1.1 & 1 \\
HD 155806  & & -2.4 & 1.2 & \\
HD 179406   & & -2.6 & 1.5 & \\
HD 151804  & & -5.7 & 1.3 & \\
HD 41117\tablenotemark{c}  & 13.72 & 16.3 & 1.2 & 1 \\
HD 152408  & & -2.3 & 1.6 & \\
HD 170740\tablenotemark{c}  & -10.83 & -5.8 & 1.1 & 2 \\
HD 43384  & & 17.1 & 1.1 & \\
HD 223385  & -22.30 & -24.2 & 2.2 & 1 \\
HD 149404  & & (1.0)\tablenotemark{b} & ...  & \\
HD 154368\tablenotemark{c}   & -5.87 & -0.8 & 1.4 & 1 \\
HD 148379  & & -5.6 & 2.0 & \\
HD 152235  & & -8.4 & 1.4 & 3 \\
HD 185247  & & -19.3 & 1.1 & \\
HD 147889\tablenotemark{c}  & -7.73 & -15.3 & 0.9 & 2 \\
HD 169454  & & 2.1 & 1.3 & \\
HD 183143  & -2.94 & 5.6 & 1.6 & 4 \\
HD 168625   & & 9.7 & 1.8 & \\
HD 168607   & & 7.9 & 1.6 & \\
Cyg OB2 No.\,8   & & -2.0 & 1.8 & \\
Cyg OB2 No.\,10   & & -0.8 & 2.1 & \\
Cyg OB2 No.\,3  & & -0.9 & 1.6 & \\
Cyg OB2 No.\,5  & -10 & (-2.5)\tablenotemark{b} & ... & 5 \\
Cyg OB2 No.\,9  & & 0.8 & 1.8 & \\
Cyg OB2 No.\,12  & -10 & 5.7 & 1.8 & 5 \\
Wd 1 W 7   & & -19.0 & 2.9  & \\
Wd 1 W 33   & & -20.8 & 2.7 & \\
\enddata
\tablecomments{}
\tablenotetext{a}{1: \citet{wel01}, 2: \citet{sie20}, 3: \citet{cra01}, 4: \citet{hob08}, 5: \citet{mcc02}}
\tablenotetext{b}{The velocities measured for DIB $\lambda 11797$ are shown instead, because the DIB $\lambda10780$ velocities could not be evaluated, owing to the broad line of a stellar \ion{He}{1} 10830 line.}
\tablenotetext{c}{\textcolor{black}{The stars are used to determine the rest-frame wavelength of DIB $\lambda 10780$, because the FWHMs of DIB $\lambda 10780$ are narrower than 1.5\AA  and the velocities of the interstellar \ion{K}{1} absorption lines are known.}}
\label{velocity}
\end{deluxetable}

\subsection{DIB search}

%
We used the reduced spectra of all the targets to search for new DIBs in the range of 0.91--1.33 $\mu$m. Our strategy for distinguishing DIBs from stellar absorption lines was to compare the heavily reddened targets to reference stars without any extinction. Initial candidate DIBs were acquired by comparing the spectra of the most heavily reddened stars ---HD183143, HD168625, HD168607, and the stars of Cyg OB2 and Wd 1--- to those of the reference star, $\beta$ Ori. 
Since the spectral types of some highly reddened stars are earlier than that of $\beta$ Ori, the strengths of several lines change from object to object. The clearest difference between late-B and early-B/O stars is the line-strength ratio of \ion{He}{2} to \ion{He}{1}. Thus, we compared the DIB candidates detected in the Cyg OB2 members, but not in HD183143 against the spectra of HD 210191 (B2III), to confirm that they were not stellar absorption lines. Although there are intervening clouds toward HD 210191 ($E(B$ - $V)=0.04$), this star can be used as a semi-standard star, owing to the weakness of the strongest DIBs in the range (e.g., $\lambda 10780 < 5$ m\r{A}). Note that the spectral resolution difference between the reference stars ($R = 28,000$ or $\Delta v = 11$ km s$^{-1}$) and a subset of the reddened stars ($R\sim 20,000$ or $\Delta v \sim 15$ km s$^{-1}$) did not affect the DIB search, because most of the DIBs and stellar lines are broader than 30 km s$^{-1}$, which is broader than the velocity widths of the instrumental profiles . 

To find new DIB candidates, we first shifted the spectra of the heavily reddened stars into alignment using the DIB $\lambda10780$ velocities (Table \ref{velocity}), and we then extracted candidates by comparing the spectra of the reddened stars with those of the reference stars. The candidate search was conducted by eye, while the spectra of telluric-standard stars were simultaneously checked, to avoid misidentification, owing to residual features from the strong telluric absorption lines. The rest-frame wavelengths of the DIB candidates were measured using the shifted spectra of the reddened stars.

After this initial search, we searched for absorption in the wavelengths of the DIB candidates toward all the targets within a range of $-30<v_\text{10780}<30$ km s$^{-1}$ using the software described in \textsection{3.2}. If any absorption bands with similar wavelengths to those of the DIB candidates were detected with EWs at the \textcolor{black}{5$\sigma$ level}, we measured the parameters of the DIBs as in \textsection{3.2}; in the case where no absorption bands were detected, we calculated the upper limits. In cases where stellar absorption lines, residuals of telluric absorption lines, or spurious features significantly affected the DIB candidates, we evaluated neither the EW nor the upper limit. After checking the measurements of the DIB candidates for all targets, the parameters of each detected DIB candidate were checked. Compared with the velocities of DIB $\lambda 10780$, the DIB candidate absorptions for which the velocities deviated from $v_\text{10780}$ at more than a 3$\sigma$ level were rejected. Additionally, an FWHM histogram was produced for each DIB candidate, and the DIB candidate absorptions that were peculiarly broad or narrow were also rejected. The DIB candidates that significantly blended with other features were also removed. In the case of a slight blending with other features, the integration range was arbitrarily adjusted. In that case, therefore, the systematic uncertainties caused by the blending should be included in the parameters of the DIB candidates.

The candidates that satisfy the following requirements are regarded as DIBs: (1) the candidates are detected toward more than 10 reddened stars; and (2) the correlation coefficients between the candidates' EWs and $E(B-V)$ are larger than zero, with $p < 0.05$. The multipeaked DIBs were treated as a single feature, because we could not determine whether these multiple peaks originated from the intrinsic profile of an identical DIB carrier or the blended profile of multiple DIBs.


\section{Results}

\subsection{Detected and candidate DIBs}


\startlongtable
\begin{deluxetable*}{cccccccc}
\tabletypesize{\scriptsize}
\tablecaption{Detected DIBs \label{catalog}}
\tablehead{
\colhead{Wavelength}	 & \colhead{Wavenumber}	 & \colhead{Detection}	  & \colhead{EW/E(B-V)} & \colhead{$R_{E(B-V)}$}	 & \colhead{FWHM} & \colhead{Reference\tablenotemark{a}} & \colhead{\textcolor{black}{Blending feature}} 	\\
\colhead{(\r{A})} & \colhead{(cm$^{-1}$)} & \colhead{} & \colhead{(m\r{A} mag$^{-1}$)} &  & \colhead{\r{A}} & \colhead{}
}
\startdata
9577.0 & 10438.8 & 30/31 & 118.0 $\pm$ 3.3  & 0.79 & 2.8 $\pm$ 0.4 & (2) &  \\
9632.1 & 10379.1 & 19/31 & 95.0 $\pm$ 2.9  & 0.84 & 2.0 $\pm$ 0.4 &  (2) & \ion{Mg}{2}\\
9673.4 & 10334.8 & 21/31 & 12.2 $\pm$ 1.1  & 0.92 & 1.0 $\pm$ 0.1 & This work &  \\
9880.1 & 10118.5 & 23/31 & 22.7 $\pm$ 1.2  & 0.72 & 1.1 $\pm$ 0.2 & (5) & \textcolor{black}{\ion{O}{2}} \\
9987.0 & 10010.3 & 14/31 & 20.4 $\pm$ 0.9  & 0.84 & 2.0 $\pm$ 0.1 & This work &  \\
10006.6 & 9990.7 & 15/31 & 16.1 $\pm$ 1.0  & 0.72 & 1.5 $\pm$ 0.4 & This work &  \\
10262.5 & 9741.5 & 12/31 & 6.0 $\pm$ 0.7  & 0.63 & 1.0 $\pm$ 0.2 & (6) &  \\
10288.0 & 9717.4 & 21/31 & 8.9 $\pm$ 0.9  & 0.57 & 1.2 $\pm$ 0.2 & This work &  \\
10360.7 & 9649.2 & 26/31 & 27.4 $\pm$ 0.9  & 0.87 & 1.7 $\pm$ 0.3 &  (4) &  \\
10393.5 & 9618.8 & 25/31 & 18.0 $\pm$ 1.2  & 0.91 & 1.0 $\pm$ 0.2 & (4) &  \\
10439.0 & 9576.8 & 26/31 & 29.4 $\pm$ 1.6  & 0.78 & 2.3 $\pm$ 0.3 & (4) &  \\
10504.4 & 9517.2 & 19/31 & 31.5 $\pm$ 1.3  & 0.94 & 1.1 $\pm$ 0.1 & (4) & \ion{Fe}{2}, \ion{N}{1}  \\
10542.6 & 9482.8 & 17/31 & 16.8 $\pm$ 1.3  & 0.76 & 1.0 $\pm$ 0.2 & This work &  \ion{N}{1}? \\
10610.3 & 9422.2 & 24/31 & 11.4 $\pm$ 0.9  & 0.78 & 1.2 $\pm$ 0.3 & This work &  \\
10697.6 & 9345.3 & 26/31 & 149.5 $\pm$ 2.2  & 0.87 & 4.3 $\pm$ 0.3 & (4) &  \\
10734.5 & 9313.2 & 22/31 & 19.1 $\pm$ 1.2  & 0.75 & 1.5 $\pm$ 0.2 & (6) & \ion{O}{1} \\
10771.9 & 9280.9 & 19/31 & 16.0 $\pm$ 1.2  & 0.87 & 1.5 $\pm$ 0.3 & This work &  \\
10780.6 & 9273.4 & 28/31 & 134.0 $\pm$ 1.5  & 0.91 & 1.2 $\pm$ 0.1 & (3) &  \\
10792.3 & 9263.3 & 25/31 & 33.9 $\pm$ 1.3  & 0.79 & 1.7 $\pm$ 0.2 & (3) &  \\
10813.9 & 9244.8 & 18/31 & 20.5 $\pm$ 1.4  & 0.70 & 1.6 $\pm$ 0.4 & This work &  \\
10876.9 & 9191.3 & 23/31 & 12.7 $\pm$ 1.1  & 0.56 & 1.6 $\pm$ 0.2 & This work &  \\
10883.9 & 9185.4 & 25/31 & 19.1 $\pm$ 1.2  & 0.82 & 1.2 $\pm$ 0.2 & (6) &  \\
10893.9 & 9176.9 & 19/31 & 15.5 $\pm$ 1.3  & 0.66 & 1.6 $\pm$ 0.3 & This work &  \\
11018.2 & 9073.4 & 11/31 & 10.6 $\pm$ 1.1  & 0.76 & 0.9 $\pm$ 0.2 & This work &  \\
11691.6 & 8550.8 & 15/31 & 17.2 $\pm$ 1.3  & 0.80 & 1.3 $\pm$ 0.2 & This work &  \\
11695.0 & 8548.3 & 22/31 & 21.0 $\pm$ 1.3  & 0.84 & 1.4 $\pm$ 0.1 & (6) &  \\
11698.5 & 8545.7 & 27/31 & 29.5 $\pm$ 1.6  & 0.92 & 1.4 $\pm$ 0.3 & (6) &   \\
11709.9 & 8537.5 & 14/31 & 9.0 $\pm$ 0.9  & 0.72 & 1.0 $\pm$ 0.2 & This work &  \\
11720.8 & 8529.5 & 27/31 & 38.7 $\pm$ 1.7  & 0.86 & 2.1 $\pm$ 0.1 & (6) &  \\
11792.5 & 8477.6 & 22/31 & 18.6 $\pm$ 1.4  & 0.85 & 1.0 $\pm$ 0.3 & (6) &  \\
11797.5 & 8474.0 & 30/31 & 119.8 $\pm$ 1.7  & 0.91 & 1.6 $\pm$ 0.1 & (1) &  \\
11863.5 & 8426.9 & 20/31 & 10.4 $\pm$ 1.2  & 0.54 & 1.3 $\pm$ 0.3 & This work &  \\
11929.3 & 8380.4 & 15/31 & 14.0 $\pm$ 0.9  & 0.77 & 1.5 $\pm$ 0.3 & This work &  \\
12194.4 & 8198.2 & 10/31 & 8.4 $\pm$ 0.8  & 0.91 & 0.9 $\pm$ 0.2 & This work &  $P(12)$, $Q(18)$ and $R(28)$ of C$_2$ (0,0)\\
12200.7 & 8194.0 & 26/31 & 17.7 $\pm$ 1.2  & 0.83 & 1.5 $\pm$ 0.4 & This work &  $P(14)$ and $R(30)$ of C$_2$ (0,0)\\
12222.5 & 8179.4 & 28/31 & 30.9 $\pm$ 1.2  & 0.90 & 1.2 $\pm$ 0.2 & (6) &  \\
12230.0 & 8174.4 & 19/31 & 14.3 $\pm$ 1.0  & 0.89 & 1.1 $\pm$ 0.2 & This work &  \\
12294.0 & 8131.8 & 29/31 & 20.5 $\pm$ 1.1  & 0.71 & 1.9 $\pm$ 0.1 & (5) & \ion{N}{1} \\
12313.5 & 8118.9 & 22/31 & 7.2 $\pm$ 0.7  & 0.45 & 1.4 $\pm$ 0.3 & This work &  \\
12337.1 & 8103.4 & 29/31 & 110.8 $\pm$ 2.0  & 0.92 & 3.5 $\pm$ 0.7 & (4) &  \\
12519.0 & 7985.7 & 20/31 & 21.1 $\pm$ 1.2  & 0.90 & 1.7 $\pm$ 0.2 & (5) &  \\
12537.0 & 7974.2 & 26/31 & 31.1 $\pm$ 1.5  & 0.88 & 1.8 $\pm$ 0.4 & (5) &  \\
12594.9 & 7937.6 & 20/31 & 10.7 $\pm$ 1.2  & 0.74 & 1.6 $\pm$ 0.2 & This work &  \\
12624.1 & 7919.2 & 29/31 & 68.5 $\pm$ 1.9  & 0.90 & 2.4 $\pm$ 0.6 & (5) &  \\
12650.0 & 7903.0 & 21/31 & 19.7 $\pm$ 1.4  & 0.89 & 1.4 $\pm$ 0.2 & This work  &  \\
12691.9 & 7876.9 & 18/31 & 19.4 $\pm$ 1.2  & 0.67 & 2.1 $\pm$ 0.3 & This work  &  \\
12798.8 & 7811.1 & 18/31 & 35.4 $\pm$ 1.2  & 0.63 & 1.1 $\pm$ 0.2 & (5) & Pa$\beta$ \\
12837.6 & 7787.5 & 14/31 & 57.9 $\pm$ 1.9  & 0.80 & 4.3 $\pm$ 0.2 & (6) & Pa$\beta$ \\
12861.5 & 7773.0 & 26/31 & 38.3 $\pm$ 1.7  & 0.85 & 2.3 $\pm$ 0.6 & (5) & Pa$\beta$ \\
12878.9 & 7762.5 & 12/31 & 11.2 $\pm$ 1.0  & 0.73 & 1.6 $\pm$ 0.4 & This work  &  \\
13021.0 & 7677.8 & 20/31 & 14.6 $\pm$ 1.0  & 0.88 & 1.8 $\pm$ 0.3 & This work  &  \\
13027.7 & 7673.9 & 24/31 & 59.4 $\pm$ 1.7  & 0.90 & 3.5 $\pm$ 0.6 & (4) &   \\
13050.5 & 7660.4 & 17/31 & 11.7 $\pm$ 1.1  & 0.82 & 1.6 $\pm$ 0.2 & This work  &  \\
13175.9 & 7587.5 & 30/31 & 442.7 $\pm$ 3.1  & 0.91 & 4.1 $\pm$ 0.3 & (1) &  \\
\enddata
\tablecomments{}
\tablenotetext{a}{\textcolor{black}{The references that confirmed the DIBs for the first time: }(1) \citet{job90}; (2) \citet{foi94}; (3) \citet{gro07}; (4) \citet{cox14}; (5) H15; and (6) \citet{ebe22}}
\label{catalog}
\end{deluxetable*}

\begin{deluxetable*}{ccccccc}
\tabletypesize{\scriptsize}
\tablecaption{Candidate DIBs}
\tablewidth{0pt}
\tablehead{
\colhead{Wavelength}	 & \colhead{Wavenumber}	 & \colhead{Detection}	  & \colhead{EW/E(B-V)} & \colhead{$R_{E(B-V)}$ ($p$-value)}	 & \colhead{FWHM} & \colhead{\textcolor{black}{Blending Feature}} 	\\
\colhead{(\r{A})} & \colhead{(cm$^{-1}$)} & \colhead{} & \colhead{(m\r{A} mag$^{-1}$)}  & \colhead{} & \colhead{\r{A}} 
}
\startdata
10280.3 & 9724.7 & 13/31 & 5.9 $\pm$ 0.7  & 0.55 (5.1e-02) & 1.0 $\pm$ 0.1 & \\
10528.3 & 9495.6 & 11/31 & 13.1 $\pm$ 1.3  & 0.10 (7.7e-01) & 1.7 $\pm$ 0.4 & \ion{Fe}{2}? \\
10650.2 & 9386.9 & 11/31 & 7.8 $\pm$ 0.9  & 0.56 (7.3e-02) & 1.0 $\pm$ 0.1 &  \ion{N}{1}? \\
12110.8 & 8254.8 & 9/31 & 26.7 $\pm$ 2.3  & 0.76 (1.8e-02) & 1.4 $\pm$ 0.2 &  $Q(8)$ of C$_2$ (0,0) \\
12175.2 & 8211.1 & 13/31 & 6.8 $\pm$ 0.9  & 0.00 (9.9e-01) & 0.9 $\pm$ 0.1 & $Q(16)$ and $R(26)$ of C$_2$ (0,0) \\
12406.2 & 8058.3 & 20/31 & 12.4 $\pm$ 1.0  & 0.20 (4.0e-01) & 1.8 $\pm$ 0.2 & \\
12486.7 & 8006.3 & 9/31 & 4.7 $\pm$ 0.7  & 0.77 (1.5e-02) & 1.1 $\pm$ 0.1 & \\
12668.3 & 7891.6 & 10/31 & 5.8 $\pm$ 0.8  & 0.60 (6.5e-02) & 1.5 $\pm$ 0.1 & \\
\enddata
\tablecomments{}
\label{catalog_cand}
\end{deluxetable*}

As a result of the search for new DIBs in \textsection{3.3}, we detected \textcolor{black}{34} new DIBs, nine of which were independently reported by \citet{ebe22}. In addition, we also detected \textcolor{black}{eight} candidates. 
Figure \ref{montage1} shows the spectra of  \textcolor{black}{54} detected DIBs, \textcolor{black}{25} of which were newly detected in this study. The spectra of two stars with no or low reddening, $\beta$ Ori and HD 210191, and the stars with the largest reddening in the Cyg OB2 association (No.\,12) and Wd 1 (W 33) are plotted. \textcolor{black}{In Figure \ref{montage1}, the DIBs detected toward W 33 have broader profiles compared to those of Cyg OB2 No.\,12, on average, probably because of the Doppler broadening.} Tables \ref{catalog} and \ref{catalog_cand} summarize the results of the measurements for the detected DIBs and candidate DIBs, respectively. Tables \ref{catalog} and \ref{catalog_cand} show the rest-frame wavelengths (column 1), wavenumbers (column 2), numbers of detections (column 3), EWs per a unit of $E(B-V)$ (column 4), the correlation coefficients with $E(B-V)$ (column 5), FWHMs in a wavelength scale (column 6), and comments on the blending features (column 7). The DIB rest-frame wavelengths were determined by minimizing the squared difference between the velocities of DIB $\lambda 10780$ (Table \ref{velocity}) and the DIB velocities, which depend on the rest-frame wavelengths of the DIBs. The EWs per a unit of $E(B-V)$ were calculated as $W/E(B-V) = \sum_i^N W_i E_{B-V, i} / \sum_i^N E_{B-V, i}^2$. The FWHMs in Tables \ref{catalog} and \ref{catalog_cand} are the average of the FWHMs of the detected DIBs with a sigma clipping. To avoid the effect of Doppler broadening in the FWHM calculation, as far as possible, the FWHMs larger than the 1$\sigma$ level were clipped, whereas the FWHMs smaller than the 2$\sigma$ level are clipped. In the following subsections, we comment on some specific DIBs and candidates. 

\subsubsection{C$_{60}^+$ DIBs}

All five absorption bands identified as the electronic band of C$_{60}^+$ are located in the wavelength range of our observations \citep{cam15, cam17}. Two main bands, $\lambda9577$ and $\lambda9632$, which were contaminated with the strong water vapor lines, could be detected toward many targets. The latter band is known to overlap with a pair of stellar \ion{Mg}{2} lines at 9631.9 and 9632.4\r{A} \citep{jen97,gal17a,wal17}. We did not try to remove the effect of the overlapping \ion{Mg}{2} lines because that is outside the scope of this study. Therefore, we note that the measured parameters of $\lambda9632$ in Table \ref{catalog} are affected by \ion{Mg}{2} lines. Three weak subbands, $\lambda9348$, $\lambda 9365$, and $\lambda 9428$, are contaminated with stronger water vapor lines. Therefore, these bands could not be detected at all in the spectra obtained using the Araki telescope, owing to the strong water vapor lines (Figure \ref{telsite}). We tried to analyze the three subbands in the NTT spectra; however, we could not detect them clearly, owing to the intrinsic weakness of the bands, the systematic uncertainties caused by the removal of the telluric lines, and the blending stellar lines. Therefore, we did not include these three bands in Table \ref{catalog} in this study. The properties of C$_{60}^+$ in the interstellar clouds and their relations with the other interstellar features, including the newly found NIR DIBs, are of great interest. An analysis dedicated to these C$_{60}^+$ bands will be conducted in the subsequent study.

\subsubsection{$\lambda11691$, $\lambda11695$, and $\lambda11698$}

Three DIBs, $\lambda11691$, $\lambda11695$, and $\lambda11698$, were detected in the wavelength range where the telluric lines are strong. Because the profiles of these DIBs are clearly resolved in the spectra of most of the targets, we have treated these features as independent DIBs in this study. However, the EWs of these DIBs are well correlated with each other ($r \sim 0.9$). Considering that these DIBs are affected by strong telluric lines, which increase the scatter in the correlation, it is possible that these DIBs originate from the same carrier, and the separations and the relative intensities of the three peaks may be important for constraining the molecular properties of the carrier molecule.

\subsubsection{DIBs overlapping with the C$_2$ Phillips bands}

The electronic bands of C$_2$ and CN are located in the range of 0.91--1.33 $\mu$m \citep{ham19}. In particular, the C$_2$ bands have many rotational lines in a wide wavelength range, because C$_2$ can be rotationally excited to higher levels. The rotational lines of the $J''=0-20$ of the C$_2$ (1,0) and (0,0) bands range between 10130--10300\r{A} and 12070--12332\r{A}, respectively. It was difficult to find weak DIBs in these wavelength ranges of the C$_2$ bands, because the C$_2$ bands were strong in the spectra of most of the stars in the Cyg OB2 association and the Wd 1 cluster, which were used for the initial search for new DIBs. Four DIBs---$\lambda$12194, $\lambda$12200, $\lambda$12294, and $\lambda$12313---and \textcolor{black}{a DIB candidate $\lambda$12110}, were detected in the region of the C$_2$ (0,0) band, while we could not find any new DIBs in the region of the C$_2$ (1,0) band. In particular, $\lambda$12110 is blended with a $Q(8)$ line, which is relatively strong among the C$_2$ lines. We did not deblend the $Q(8)$ line from $\lambda$12110 by simulating the C$_2$ line profile. Therefore, $\lambda$12110 was measured only in the spectra of the stars toward which the C$_2$ band was not detected or very weak.

\subsubsection{DIBs detected in the wing of Pa $\beta$}

DIBs $\lambda$12799, $\lambda$12838, and $\lambda$12862 were detected on the wing of the broad Pa$\beta$ absorption line. We removed this broad wing by fitting a polynomial function for measuring the EWs of these DIBs. 

\begin{figure*}
\includegraphics[width=18cm,clip]{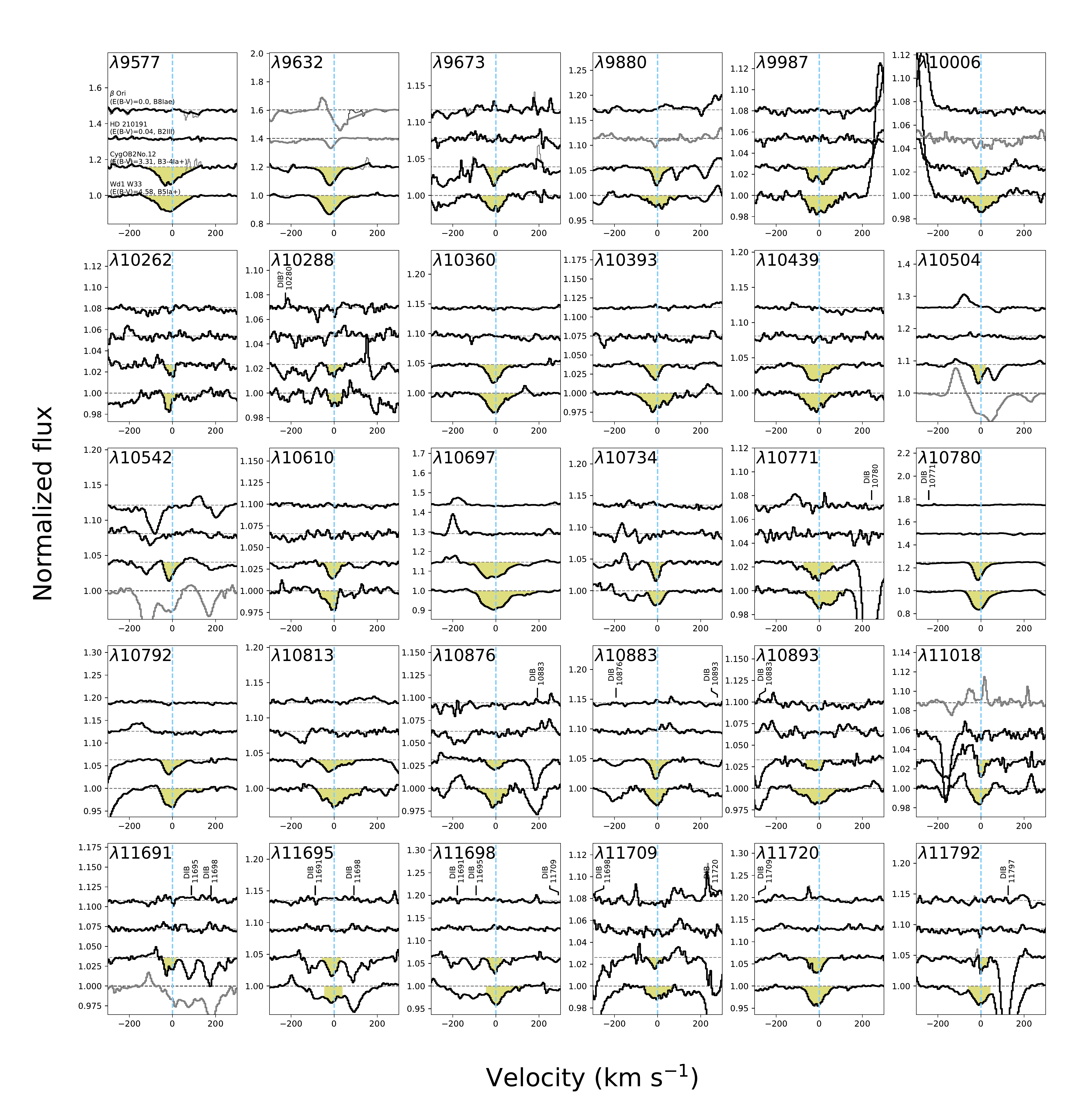}
\caption{The spectra of all  \textcolor{black}{54}  DIBs detected in the 0.91--1.33 $\mu$m range. The spectra of $\beta$ Ori (the top spectrum), HD210191 (the second spectrum from the top), Cyg OB2 No.\,12 (the third spectrum from the top), and Wd1 W33 (the bottom spectrum) are shown in each panel. The horizontal axis shows the line-of-sight velocities that were calculated with the rest-frame wavelengths shown in Table \ref{catalog}. The spectra are shifted by the velocities that were measured using the DIB $\lambda$10780, to align the DIBs. The regions that are contaminated with the strong telluric absorption lines are interpolated from the surrounding regions, which the strong telluric absorption lines do not contaminate. The interpolated and original spectra are shown using the thick and thin lines, respectively. The spectra are shown with gray lines in cases where the DIBs could not be evaluated, owing to the blending of the other features. The detected DIBs are indicated by the yellow areas. \textcolor{black}{The marks with the black thin lines that are sometimes plotted above the spectra show DIBs and candidates that appear in the panels of the other DIBs.} 
}
\label{montage1}
\end{figure*}

\begin{figure*}
\figurenum{4}
\includegraphics[width=18cm,clip]{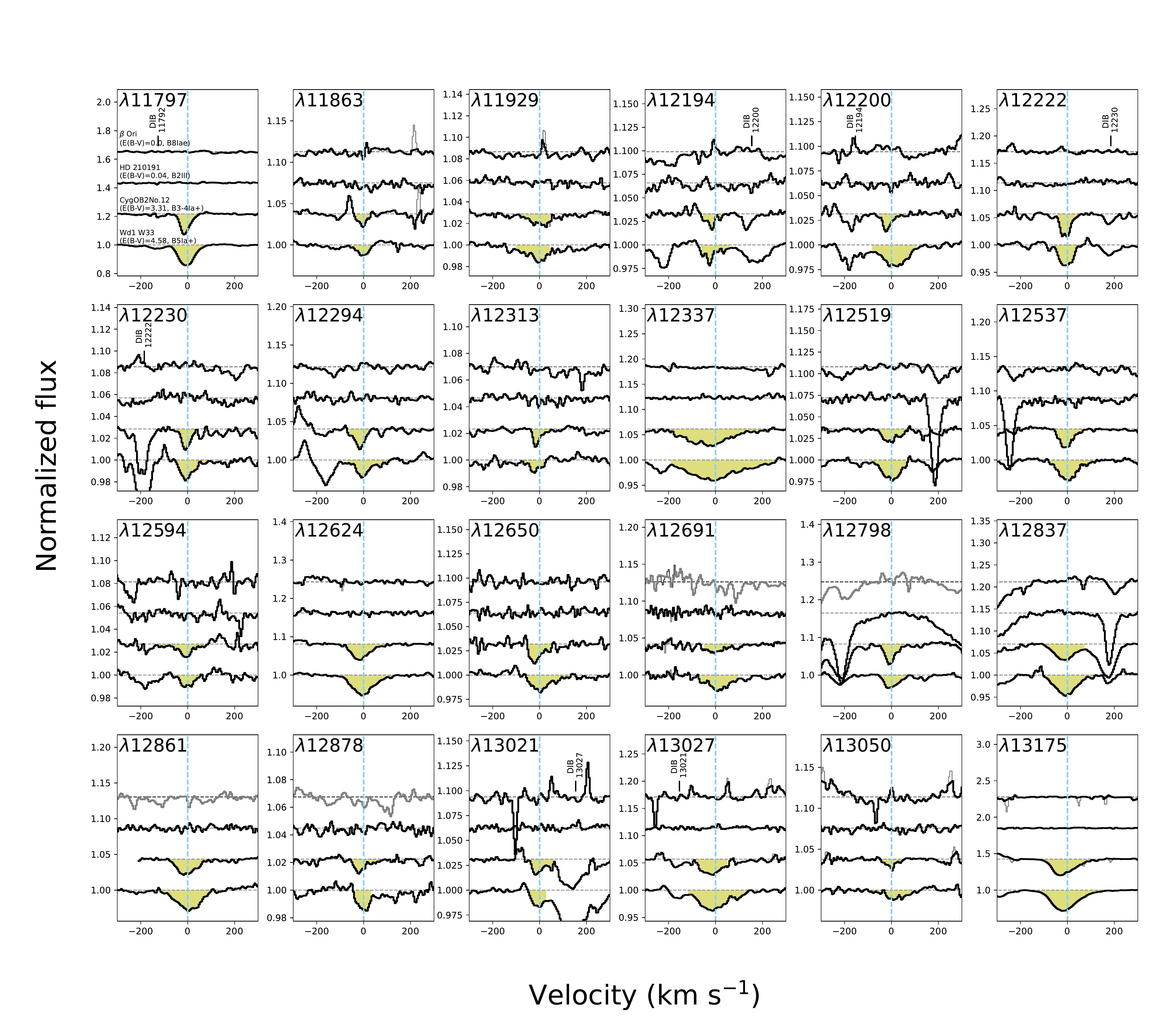}
\caption{\textit{Continued.}}
\label{montage2}
\end{figure*}

\subsection{NIR DIBs in the 0.91--1.33 $\mu$m range}

\begin{figure}
\includegraphics[width=9cm,clip]{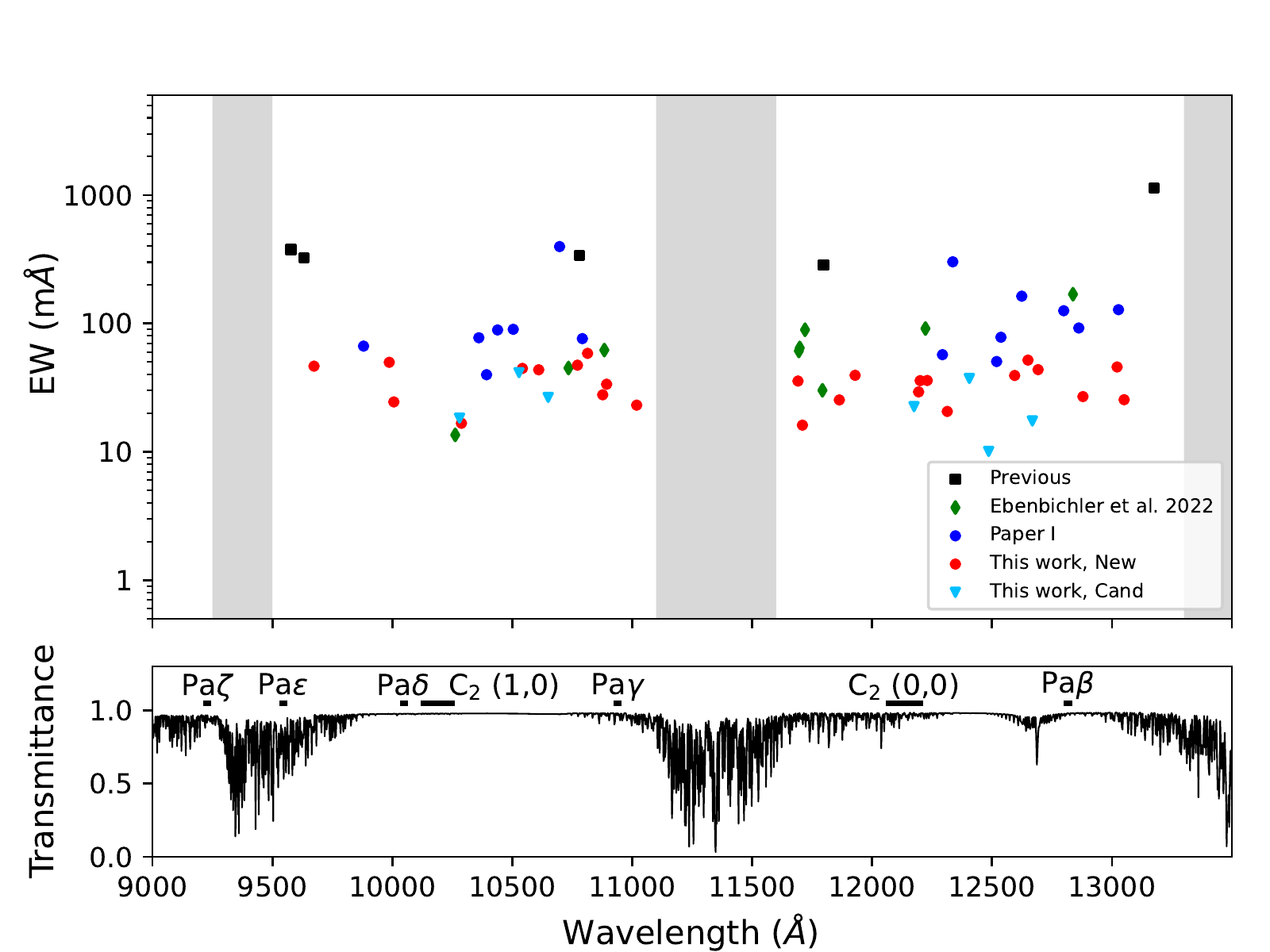}
\caption{Upper panel: the EWs of the DIBs detected toward No.\,12 in the range of 0.91--1.33 $\mu$m. The circles show the DIBs detected or confirmed in H15 (blue), and the DIBs detected in this study (red). The black squares show the DIBs detected in the literature \citep{job90,foi94,gro07}. The green diamonds show the DIBs detected in \citet{ebe22}. The light blue triangles show the candidate DIBs in this study. 
Lower panel: transmittance spectrum of the atmosphere. The spectrum is synthesized using ATRAN. The locations of the Paschen \ion{H}{1} lines and C$_2$ Phillips bands are also shown. }
\label{ewdNo12}
\end{figure}

\begin{figure}
\includegraphics[width=9cm,clip]{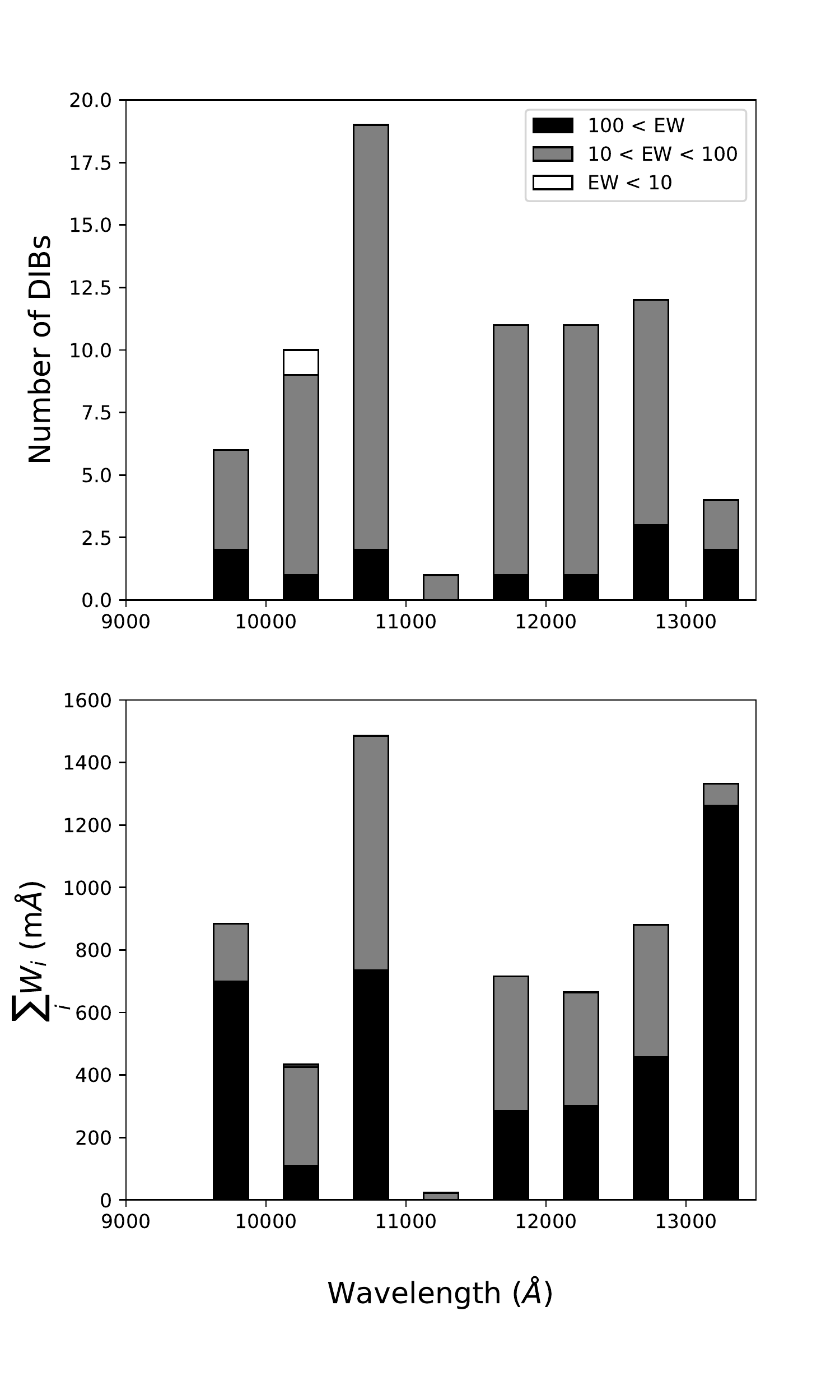}
\caption{Number density (top) and EW density (bottom) distributions toward Cyg OB2 No.\,12. The bin width is 500 \r{A} in all panels. The black, gray, and white bars show DIBs with EW $>$ 100 m\r{A}, 10 $<$ EW $<$ 100 m\r{A}, and EW $<$ 10 m\r{A}, respectively.} 
\label{histNo12}
\end{figure}

Figure \ref{ewdNo12} shows the EWs of the NIR DIBs in the range of 0.91--1.33 $\mu$m for Cyg OB2 No.\,12, toward which all \textcolor{black}{54} DIBs were detected, except for DIB $\lambda$12110, which was blended with the C$_2$ band. 
The EWs of 20 of the DIBs in this object that were detected in previous studies \citep[H15, ][]{cox14, job90, foi94, gro07} are larger than 30 m\r{A}, whereas the EWs of the \textcolor{black}{weakest} DIBs that are newly detected in this study are as small as 10 m\r{A}. The lack of DIBs in $\lambda < 9500$\r{A}, $11100 < \lambda < 11500$\r{A}, and $\lambda > 13300$\r{A} in Fig. \ref{ewdNo12} is a result of the strong telluric absorption lines. Accordingly, there are fewer DIBs in and near these wavelength ranges in Fig. \ref{histNo12}, which shows the distribution of the DIB numbers and EWs for Cyg OB2 No.\,12. The peaks of the number density distributions in the $Y$ and $J$ bands are at approximately 10500 and 12500 \r{A}, respectively, where the telluric absorption lines are very weak. The weakest DIBs in the $Y$ and $J$ bands are also detected in these bins. These results suggest that the detection sensitivity of the NIR DIBs is strongly influenced by the accuracy of the telluric absorption. 

\begin{figure*}
\includegraphics[width=18cm,clip]{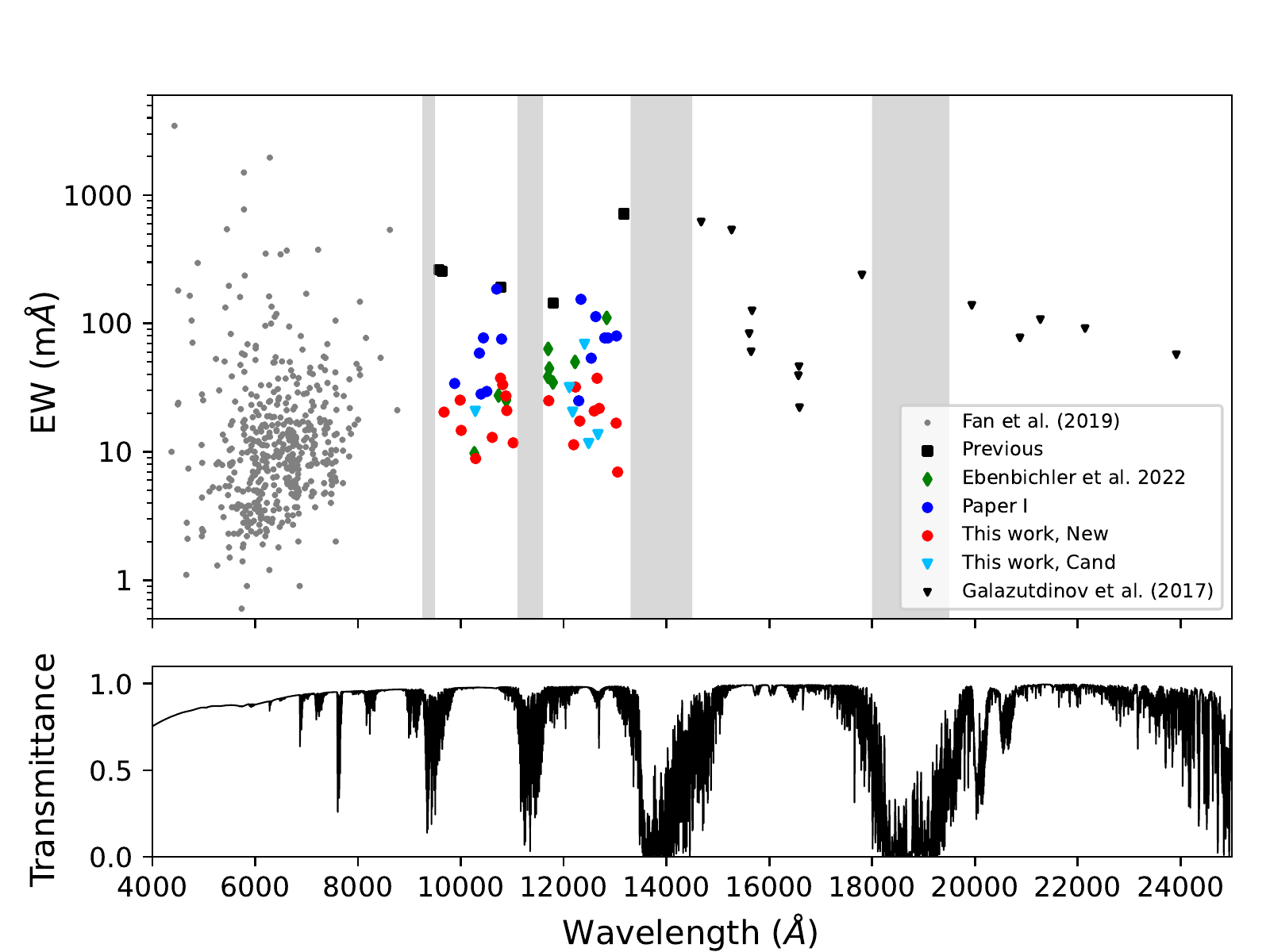}
\caption{Upper panel: overview of DIBs from the optical wavelength range to the $K$ band. The circles show the DIBs detected or confirmed in H15 (blue), and the DIBs detected in this study (red). The black squares show the DIBs detected in the literature \citep{job90,foi94,gro07}. The green diamonds show the DIBs detected in \citet{ebe22}. The light blue triangles show the candidate DIBs in this study. 
Gray and black circles show the DIBs in the range of 4000--9000 \r{A} \citep{fan19} and the $H$-band DIBs \citep{cox14}, respectively. For the DIBs in the $K$ band and some weak DIBs in the $H$ band, we adopt the DIB EWs for Cyg OB2 No.\,5 from \citet{gal17b}, because, to date, these DIBs have not been detected toward HD183143. Note that the points of No.\,5 are only plotted to show the DIB distribution at the longest wavelength, and cannot be directly compared with the other data from HD183143, owing to the difference in the DIB EWs between HD183143 and No.\,5. The gray shaded areas show the wavelength ranges in which we could not search for DIBs, owing to strong telluric absorption. Lower panel: synthetic spectrum of telluric absorption created by ATRAN \citep{lor92}.}
\label{ewdHD183143}
\end{figure*}

Figure \ref{ewdHD183143} shows the overall distribution of the central wavelengths of the DIBs, from the optical wavelength range to the $K$ band, where the longest-wavelength DIBs have been detected. This figure primarily uses the DIB EWs of HD183143 obtained by \citet[][for 4000--9000 \r{A}]{fan19}, those obtained in this study (for 9100--13300 \r{A}) and those from \citet[][for 15000--18000 \r{A}]{cox14}. For the DIBs in the $K$ band and some weak DIBs in the $H$ band, we adopt the DIB EWs for Cyg OB2 No.\,5 from \citet{gal17b}, because, to date, these DIBs have not been detected toward HD183143. Although the number of DIBs detected in the NIR region has gradually increased over the course of the past decade, previous surveys could not detect DIBs in the range of $W<$ 50 m\r{A}, in which the majority of the optical DIBs are distributed. 
In this study, we were able to detect a number of weak NIR DIBs, which has enabled us to discuss the DIB distribution over a wide wavelength range (see Section 5). It will be possible to investigate the profiles of these DIBs, as well as their correlations with gaseous parameters (e.g., \ion{H}{1} column density), and to use them for comparison with the laboratory spectra of candidate molecules for new NIR DIBs. These new DIBs will provide new insights into the mysterious carriers of DIBs. In particular, in our forthcoming study, we will investigate the DIB--$E(B-V)$ and DIB--DIB correlations.

\section{Discussion}

\subsection{Distributions of DIB numbers and strength}

In this subsection, we discuss the distributions of the DIB numbers and strength, using the DIBs of HD183143, which is most frequently used for the DIB surveys. The DIBs toward HD183143 have been searched for from the optical wavelength range to the NIR range, at high sensitivities. 
In Section 5.1.1, we discuss the DIB distributions over a wide wavelength range, from the optical wavelength range to the $H$ band. In Section 5.1.2, we attempt to deduce the intrinsic distribution of the DIBs toward HD183143 by considering the biases among the observations in different wavelength ranges, such as the differences in S/N, the spectral resolutions, and telluric absorption lines.

\subsubsection{DIBs in the 0.4-1.7 $\mu$m range}

Herein, we discuss the number and EW distributions for all the DIBs ever detected toward HD183143, to investigate the wavelength ranges (and, thus, the energies of the transitions) in which the DIBs are the most populous. 
From the previous surveys in the optical wavelength range, we have adopted the most extensive catalog of DIBs, which was reported by \citet{fan19}, who obtained a high-sensitivity (S/N $\sim 1300$ at the median) and high-resolution ($R=38,000$) spectrum of HD183143 and detected 472 bands between 4000 and 9000 \r{A}. The EWs of the DIBs of HD183143 in the $H$ band (15,000 $<\lambda<$ 18,000 \r{A}) are adopted from \citet{cox14}. 

\begin{figure}
\includegraphics[width=9cm,clip]{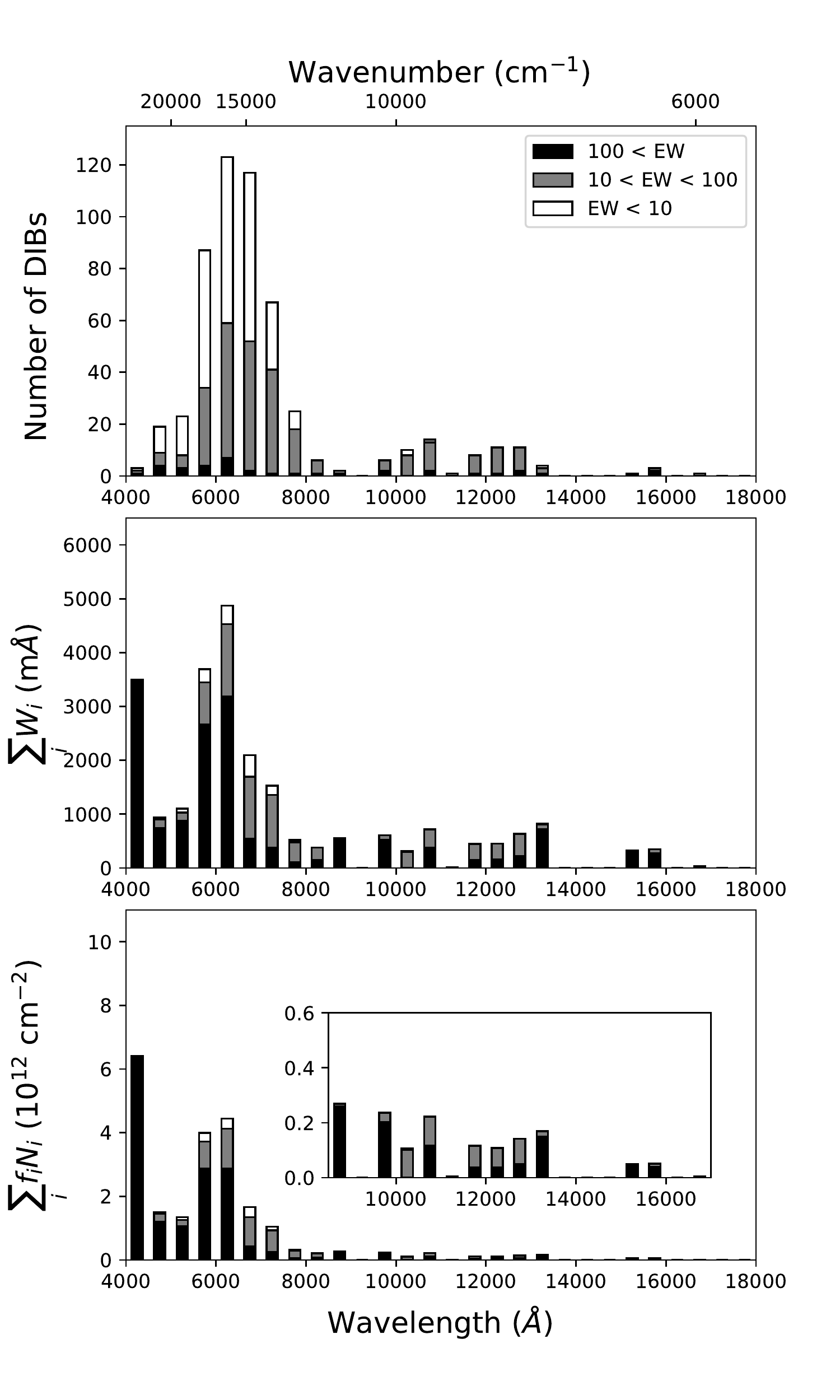}
\caption{Distributions of the numbers (upper panel), total EWs (middle panel), and $\sum_i f_i N_i$ (lower panel) of the DIBs. The horizontal axes indicate the wavelength, and the bin size is 500 \r{A}. The inset panel in the lower panel shows a detailed view of the histograms at $\lambda > 7500$ \r{A}. The colors indicate the same factors as in Fig. \ref{histNo12}.}
\label{histHD183143_lam}
\end{figure}

Figure \ref{histHD183143_lam} shows the distributions of number density (top), total EW (middle), and the summation of the product of the oscillator strengths and column densities of the DIB carriers (bottom). The summation is calculated as follows: 
\begin{equation}
\sum_{i} f_i N_i (\text{cm}^{-2}) = \sum_{i} 1.13 \times 10^{20} \frac{W_i (\text{\AA})}{\lambda _i^2 (\text{\AA})}  , 
\end{equation}
where $i$ is the DIB index within each wavelength bin, $f_i$ and $N_i$ are the oscillator strength and column density of the $i$th DIB carrier, respectively, and $W_i$ and $\lambda _i$ are the EW and central wavelength of the $i$th DIB, respectively. In this calculation, an optically thin condition is assumed. 

The peak of the number density distribution (Fig. \ref{histHD183143_lam}, top panel) was around the bins from 6000 to 7000 \r{A}. According to \citet{fan19}, this peak does not reflect the intrinsic DIB distribution, because a significant fraction of the weak DIBs at $\lambda > 7000$ \r{A} were not detectable, owing to contamination by strong telluric absorption. 
The number density of the NIR DIBs at 9000 $<\lambda<$ 13500 \r{A} is considerably lower than that of the optical DIBs, which may be a result of both the intrinsic DIB distributions and the differing spectral qualities in terms of S/N, spectral resolution, and telluric lines. In particular, very few DIBs with $W < 10$ m\r{A} are detected in the NIR wavelength range. 

The difference in the detection limits among the observations has a lower impact on the EW distribution than on the number density distribution because most of the EW contribution comes from strong DIBs ($W>100$ m\r{A}). 
For example, the 4000--4500 \r{A} bin has a very high value, owing to $\lambda$4430, which is the strongest DIB ever found. Although this stochasticity primarily affects the distribution, a trend of decreased EWs at longer wavelengths is seen, which may be attributable to an intrinsic distribution of DIB EWs rather than to observational bias. 
This trend is more clearly seen in the distribution of $\sum_{i} f_i N_i$ (the bottom panel of Fig. \ref{histHD183143_lam}). 

\begin{figure}
\includegraphics[width=9cm,clip]{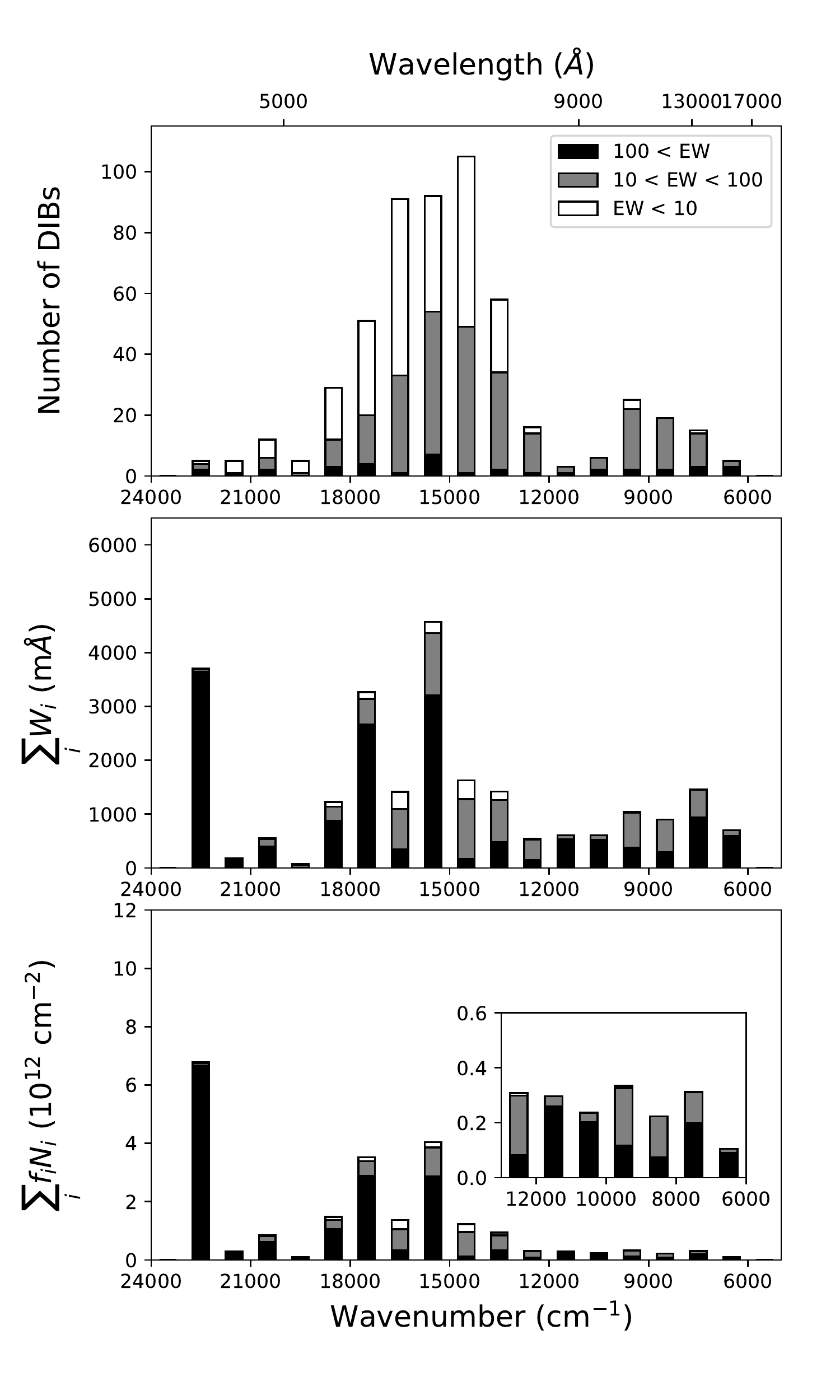}
\caption{Distributions of the numbers (upper panel), total EWs (middle panel), and $\sum_i f_i N_i$ (lower panel) of the DIBs. The horizontal axes indicate the wavenumber, and the bin size is 1000 cm$^{-1}$. The inset in the lower panel shows a detailed view of the histograms at $\nu < 13000$ cm$^{-1}$. The colors indicate the same EW ranges as in Fig. \ref{histNo12}.}
\label{histHD183143_wn}
\end{figure}

Figure \ref{histHD183143_wn} shows the DIB distributions on a wavenumber scale. The wavelength coverage of our observations corresponds to 7400 $<\nu<$ 11100 cm$^{-1}$. The wavenumber distribution is probably more appropriate than the wavelength distribution for investigating the intrinsic distributions of the DIB numbers and strengths than the wavelength distribution because the transition energies are simply proportional to the wavenumbers. The trends seen in Fig. \ref{histHD183143_lam} can also be seen in these distributions, but the differences between the optical and NIR ranges are diluted, because a bin in the wavenumber units covers a larger wavelength range at a longer wavelength. In particular, the EWs are nearly constant across the wavenumber plot, except for a few bins that contain the strongest DIBs. To understand the intrinsic distribution of these DIB properties, it is necessary to consider the effects on the distributions that arise from differences in observational conditions.

\subsubsection{Observational biases}

Next, we deduced the intrinsic distribution of the DIBs from the observed distribution by minimizing the effects of observational bias. The observed distributions were affected by observational biases on the strength and density of the telluric absorption and the detection limits. 

To remove the effect of telluric absorption, we selected five wavelength ranges in which the telluric absorption lines are weak: $5300<\lambda<6850$ \r{A}, $7300<\lambda<7590$ \r{A}, $10300<\lambda<10900$ \r{A}, $12200<\lambda<12550$ \r{A}, and $15400<\lambda<15680$ \r{A}. In the following calculation, the first wavelength range is divided into five sub-ranges: $5300<\lambda<5600$ \r{A}, $5600<\lambda<5900$ \r{A}, $5900<\lambda<6200$ \r{A}, $6200<\lambda<6500$ \r{A} and $6600<\lambda<6850$ \r{A}; the $6500<\lambda<6600$ \r{A} range was excluded to avoid the stellar H$\alpha$ line. 

For the detection limit, we used only those DIBs that were deeper than a fixed peak depth ($d_p$) threshold for all wavelength ranges, to minimize the effects of different detection limits. Since the S/N ratio and spectral resolution of the optical spectrum of HD183143 obtained by \citet{fan19} are both higher than those of our NIR spectrum of HD183143, the threshold from the NIR spectrum was set to be $d_p = 7\times 10^{-3}$, which is the 5$\sigma$ detection limit for our NIR spectrum of HD183143. 

\begin{figure}
\includegraphics[width=9cm,clip]{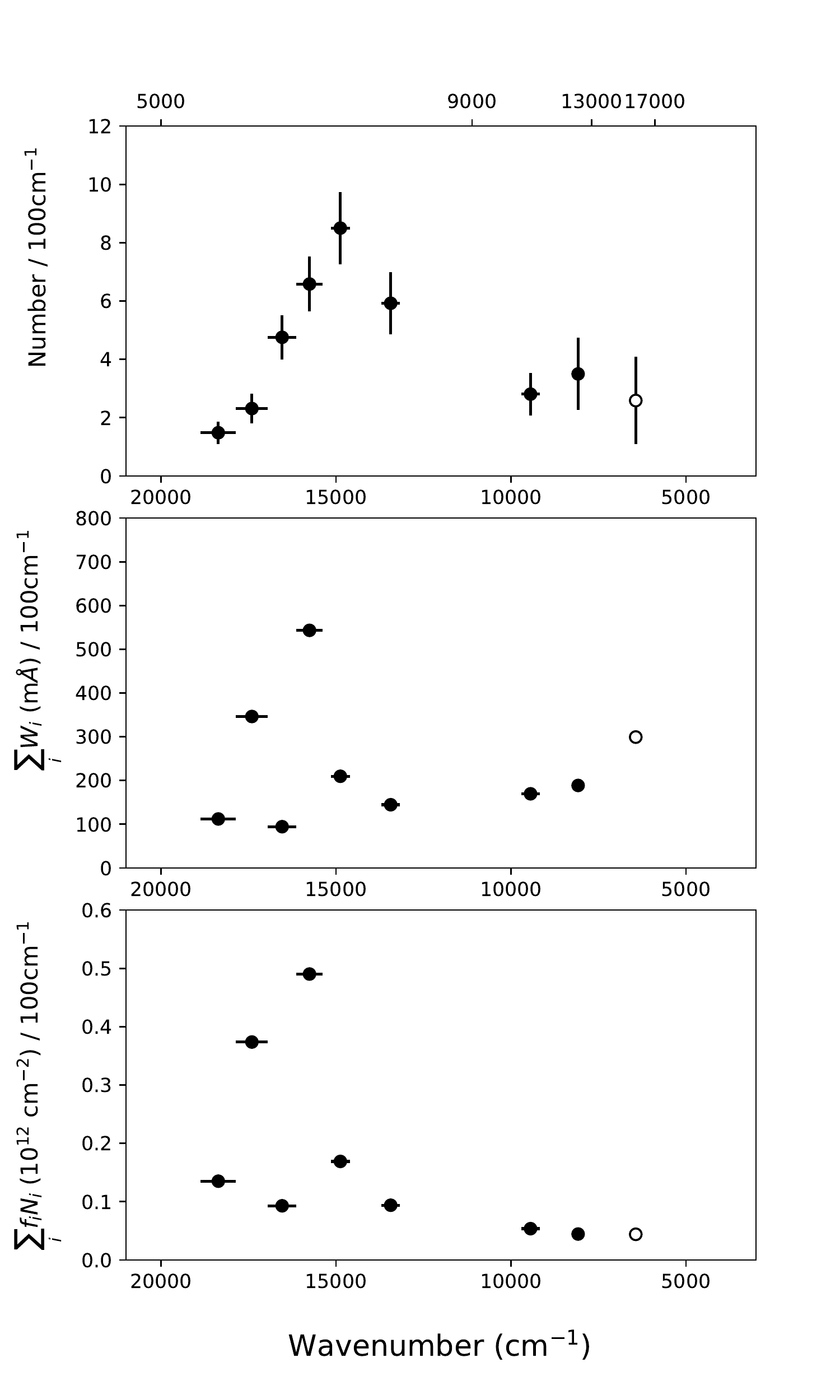}
\caption{Wavenumber densities of the DIB numbers (top panel), EWs (middle panel), and the sums of the products of the oscillator strength and column density (lower panel), after considering the observational biases. See the main text for the sampled wavelength region and the threshold used for the depth of the DIBs. The horizontal bars indicate the sampled ranges. The vertical bars in the top panel show the square roots of the numbers of DIBs, that is, the Poisson uncertainties of the DIB numbers. As a reference, we also plot the data for the $H$-band DIBs with an open circles, to see their trend at longer wavelengths. Note that the effects of the observational biases were not considered for these points. }
\label{statsDIBs}
\end{figure}

Figure \ref{statsDIBs} shows the resultant wavenumber density of the numbers, EWs, and $\sum_i f_i N_i$ of the DIBs in the sampled wavelength ranges. The number density exhibits a peak in the $6600<\lambda<6850$ \r{A} bin, and a declining trend is seen toward both the shorter- and longer-wavelength ranges. This is consistent with the results shown in Fig. \ref{histHD183143_wn}, and the peak wavelength corresponds to the 1.8 eV transition energy. In contrast, the EW density is approximately constant over the entire range, with the exception of the $6200<\lambda<6500$ \r{A} bin, in which the second-strongest DIB, $\lambda$6284.3, is located. 
The almost constant wavenumber density of the DIB EWs results in a rapid decrease of 
$\sum_{i} f_i N_i$ toward the longer-wavelength limit, as seen in the bottom panel of Figure \ref{statsDIBs}. 
Therefore, the carrier abundance of the DIBs tends to be reduced at longer wavelengths, suggesting that the NIR DIBs trace molecular species that are less abundant in interstellar clouds and/or have much smaller oscillator strengths than those in the optical region. 



\subsection{FWHM distribution}

\textcolor{black}{In this subsection, we discuss the distributions of the FWHMs of the optical and NIR DIBs for HD147889, using the data of \citet{fan19} and this study. HD147889, which is considered to be a single-cloud sightline \citep{sie20}, is a good target for investigating the DIB profiles. Here, we compare the FWHMs of the DIBs in the optical range and those in the NIR range in order to examine the relation between the DIB width and the wavelength. The intrinsic profiles of the DIBs are important for constraining the DIB carriers \citep[e.g.,][]{hua15}. If an intrinsic DIB profile reflects the rotational contour of a molecular electronic transition, the FWHM on a wavenumber scale is a function of the rotational temperature ($T_\text{rot}$) and the rotational constant ($B''$) \citep{cam04}. A difference in the FWHM distribution between the optical and NIR wavelength ranges would imply that the carriers producing the DIBs in each wavelength range have different molecular properties (e.g., electric dipole moment and molecular size), on average.}

\begin{figure*}
\includegraphics[width=18cm,clip]{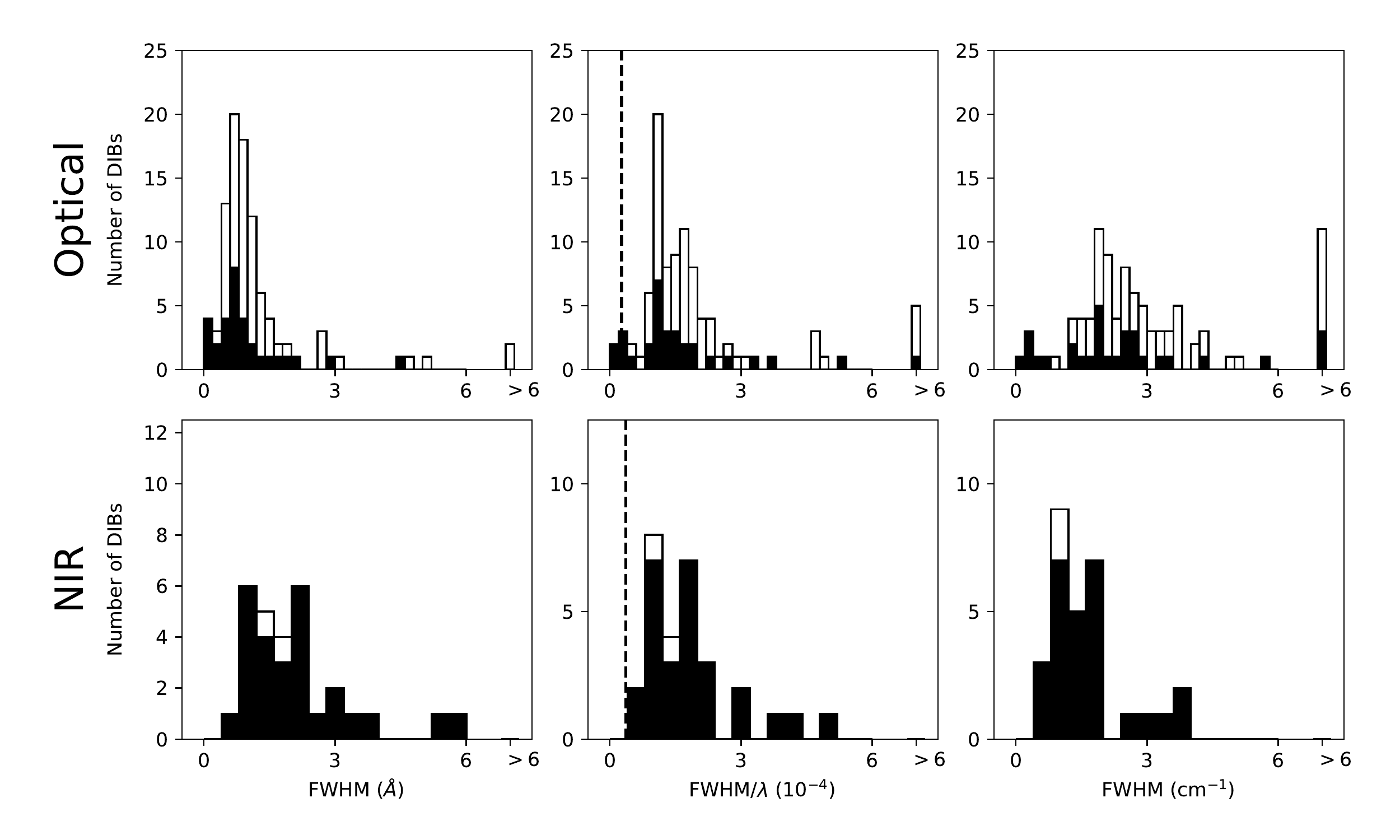}
\caption{FWHM distribution of the DIBs for HD147889. The upper and lower panels show the DIBs in the optical (5000 $<\lambda<$ 6500 \r{A}) and NIR (9100 $<\lambda<$ 13300 \r{A}) wavelength ranges, respectively. The left, middle, and right panels show the FWHMs measured in wavelengths, normalized by wavelength, and measured in wavenumbers, respectively. The dashed lines show the FWHMs corresponding to the spectral resolutions. The black bars show the DIBs with the central depths larger than 0.009.}
\label{FWHMhist}
\end{figure*}

Figure \ref{FWHMhist} shows the histograms of the DIB FWHMs measured in the wavelengths, normalized by wavelength, and measured in wavenumbers for optical DIBs \citep[$5000< \lambda <6500$ \r{A};][]{fan19} and NIR DIBs ($9100< \lambda <13300$ \r{A}; this study). The bin size for the NIR DIBs is twice as large as that for the optical DIBs, owing to the difference in their numbers. \textcolor{black}{The spectral resolutions in \citet{fan19} and in this study were 8 and 11 km s$^{-1}$, respectively. The thresholds corresponding to the spectral resolutions were also plotted with the histogram of the normalized DIB FWHMs. In the FWHM distribution on a wavenumber scale, the DIB width is narrower in the NIR range, on average. Because most of the DIBs have FWHMs broader than the spectral resolutions (see the center panels in Figure \ref{FWHMhist}), the difference of the FWHM distribution on a wavenumber scale is probably not caused by the difference of the spectral resolutions.}


The detection limits of the observations also affect the FWHM distributions. Since the broader DIBs have shallower depths, it is relatively difficult to detect broader DIBs at the same EW. Therefore, the detected weak DIBs tend to have narrower widths, and the detection limit can change the FWHM distribution. The FWHM distributions of the DIBs with a central depth larger than \textcolor{black}{0.009} are also shown with the black bars in Fig. \ref{FWHMhist}. The threshold is sufficiently higher than the detection limits of both observations, without a significant decrease in the DIB numbers. The FWHM distributions of the DIB samples that are limited by depth would reflect the intrinsic DIB properties better than that of a full sample would. Other factors of the observations, such as the contamination of the telluric absorption lines and the wavelength coverages of each echelle order, can bias the FWHM distributions; however, it is difficult to evaluate their effects.

In these depth-limited samples, the difference between the wavenumber-scale FWHM distributions is clear: the medians are \textcolor{black}{2.1 and 1.5} cm$^{-1}$ for the optical and NIR wavelength ranges, respectively. To investigate the statistical significance of the difference in the FWHM distributions, we conducted a Mann--Whitney $U$ test, which is a nonparametric statistical test with the null hypothesis that the medians of two samples are the same, under the assumption that the shapes of the distributions are identical. The sample sizes of the optical and NIR DIBs are \textcolor{black}{29 and 27}, respectively, which are sufficient for the statistical test. We rejected the null hypothesis, with a $p$-value of \textcolor{black}{0.038}, which suggests that the difference in the FWHM distributions between the optical and NIR DIBs is statistically significant. 
The difference in the FWHM distributions between the optical and NIR wavelength ranges may imply that the DIB width on a wavenumber scale becomes intrinsically narrower at longer wavelengths.

\subsection{Carriers}

\subsubsection{General implications}

Herein, we comment on the implications for the carriers of the NIR DIBs, based on our results. 
In Section 5.2, we suggested that the intrinsic FWHMs of the NIR DIBs in units of wavenumber are, on an average, narrower than those of the optical DIBs. If we assume that 
the intrinsic widths of the DIBs are determined by the rotational constants of the DIB carriers, the NIR DIB carriers should have smaller moments of inertia (i.e., they should be larger in size) than the optical DIB carriers. This is consistent with the relationship between the transition wavelengths and the molecule sizes of conjugated molecules, such as carbon chains, PAHs, and fullerenes. 
This proportional relation between the transition wavelength and the number of carbon atoms in molecules has been experimentally confirmed for the $\pi-\pi$ electronic transitions of polyacetylene chains \citep{mai98}. The relation is also seen for the PAH molecules and ions in the calculation of the wavelengths of the PAH bands that was conducted in \citet{rui05}.  

In Section 5.1, we suggested that the summation of 
the DIB column density ($\sum _i f_i N_i$) decreases with wavelength. 
We cannot determine whether oscillator strength or column density is the primary contributor to this decrease. If we assume that conjugated molecules are a potential carrier, then their oscillator strengths will be in proportion to the total electron number, i.e., the number of carbon atoms \citep[e.g., ][]{hal03}. If the carriers of the DIBs at longer wavelengths are larger in size, as suggested by their FWHM distribution, then the DIBs at longer wavelengths can be considered to have higher oscillator strengths. 
In this case, column densities associated with longer-wavelength DIBs decrease more rapidly than is observed for $\sum _i f_i N_i$ with increasing wavelength. It follows that larger DIB carriers have lower column densities in interstellar clouds, because large molecules can have a variety of structural patterns and more atoms are necessary to form bigger molecules. In summary, we suggest that DIBs at longer wavelengths tend to be caused by larger molecules.

\citet{rui05} simulated DIB spectra toward a line-of-sight cloud of HD147889, the physical parameters of which have been well constrained, under the assumption that the DIB absorption in this region is caused by interstellar PAHs. Their results suggest that larger PAH ions tend to have transitions at longer wavelengths, and that the ionization fraction of PAHs strongly depends on their size and structure. Metallicity is another key parameter that determines the DIB distribution \citep{cox06}. Further high-sensitivity observations in both the optical and NIR wavelength ranges toward various lines of sight would put constraints on the DIB carrier properties.

\subsubsection{Possible carriers}

DIBs are considered to originate from interstellar carbonaceous molecules \citep{sar08}. 
Specific candidates include carbon chain molecules, PAHs, and fullerenes. 
The only positively identified carrier to date is ionized buckminsterfullerene (C$_{60}^+$), which has been identified as the carrier of five DIBs at approximately 0.95 $\mu$m \citep{cam15,wal15,cam16a}. Other fullerenes, such as C$_{70}^{+}$ and C$_{70}^{2+}$, were also tested by the same authors \citep{cam16a,cam17}, but their transitions were not detected in the astronomical spectra. 
\citet{omo16} reviewed the properties of other fullerenes and fullerene derivatives to explore their ability to produce interstellar DIB features. Fullerene compounds may be candidates for DIB carriers. 
\citet{tom05} obtained NIR spectra of C$_{60}^-$ in its gas phase at room temperature (300K) and identified three absorption bands at 9382, 9145, and 10460 cm$^{-1}$, which corresponded to air wavelengths of 10655, 10931, and 9557 \r{A}, respectively. Near the strongest of the three bands (10655 \r{A}), we detected a weak DIB \textcolor{black}{candidate} at 10,650 \r{A} (Table \ref{catalog_cand}). The EW and FWHM of the $\lambda10650$ for Cyg OB2 No.\,12 are \textcolor{black}{27$\pm$2 m\r{A} and 2.5 \r{A}}, respectively. 
We could not detect DIBs close to the other two weaker bands at 10931 and 9557 \r{A}. The band at 10931 \r{A} overlaps the \ion{H}{1} line, and the band at 9557 \r{A} is heavily contaminated by telluric lines; therefore, we could not even set upper limits. Despite the wavelength difference between the main band at 10655 \r{A} and $\lambda$10650, a small shift such as this may be induced because the absorption band was obtained at room temperature \citep{tom05}, which can broaden and shift bands. Based on the oscillator strength of $f=0.022$ \citep{str15}, the EW toward Cyg OB2 No.\,12 (27 m\r{A}) corresponds to \textcolor{black}{$N(\text{C}_{60}^-) =1.2 \times 10^{12}$ cm$^{-2}$}. Considering the amount of C$_{60}^+$ in the line of sight of Cyg OB2 No.\,12 ($N(\text{C}_{60}^+) = 2.5 \times 10^{13}$ cm$^{-2}$ estimated from the $\lambda$9577 band), C$_{60}^-$ may be present, and it would favor denser clouds with reduced UV intensity. Although it remains a matter of speculation, DIB $\lambda$10650 is a potential candidate for the absorption band of C$_{60}^-$. Firm identification requires (1) obtaining the gas-phase spectrum of C$_{60}^-$ at a low temperature, and (2) detecting the other two subbands. Since the detection of the band at 10931 \r{A} is nearly impossible, owing to its overlap with the stellar \ion{H}{1} line, detecting the band at 9557 \r{A} would be more significant. 

Ionized PAHs are another class of DIB carrier candidates in the NIR range. 
\citet{mat05} obtained the NIR (0.7--2.5 $\mu$m) spectra of cations and anions of 27 PAHs (the largest of which was C$_{50}$H$_{22}$) using matrix isolation spectroscopy. The Ar matrix that they used can broaden and shift the obtained absorption band. Therefore, it was difficult for them to identify DIBs as the absorption bands of ionized PAHs. They demonstrated that strong and narrow bands exist in the NIR absorption spectra of PAH cations and anions. In H15 and in this study, we have detected DIBs whose wavelengths are close to the absorption bands of PAH cations and anions. If the gas-phase spectra of such ionized PAHs can be obtained, the DIBs detected here can potentially be confirmed.

\section{Summary}

We explored weak NIR DIBs the in 0.91--1.33 $\mu$m range using the NIR high-resolution ($R=20,000$ and 28,000) spectra of 31 reddened stars that were collected using the WINERED spectrograph. Our findings are summarized as follows:

\begin{enumerate}

\item The large DIB EWs toward the heavily reddened lines of sight enabled us to detect \textcolor{black}{54} DIBs, \textcolor{black}{25} of which were newly discovered by our observations. We also independently detected nine of 12 DIBs newly detected in \citet{ebe22}. We succeeded in detecting DIBs as weak as 10 m\r{A} in the NIR range. We found that, as in the optical range, many weak DIBs populate in the NIR range of 0.91--1.33 $\mu$m. The wavelength range of 0.91--1.33 $\mu$m that has been explored in this study is of great importance for the study of interstellar molecules, because it contains the absorption bands of both small and large carbon molecules, including C$_{60}^+$, C$_2$, and CN, and many anonymous DIBs.

\item The FWHMs of the NIR DIBs were found to be narrower than those of the optical DIBs, on average. Assuming that the DIB width is determined by the rotational constant, this difference suggests that the DIBs at longer wavelengths tend to be caused by larger molecules, which is consistent with the properties of conjugated molecules.

\item We investigated the distributions of the DIB numbers, EWs, and column densities, according to wavelength (wavenumber), from the optical to the NIR range. The number density of the DIBs peaks at $\lambda \sim 6600$ \r{A} and declines toward both the shorter and longer wavelengths. Additionally, the sum of the DIB column densities decreases with increasing wavelength. Assuming that the DIBs at longer wavelengths tend to originate from larger molecules, as suggested by their FWHM distributions, the oscillator strength can be assumed to be larger for DIBs at longer wavelengths. Thus, we suggest that DIBs at longer wavelengths trace less abundant molecules. 

\item The comparison of the DIB catalog that has been compiled in this study with the gas-phase spectra of candidate molecules, such as fullerenes, PAHs, and carbon chains, will contribute to the further identification of DIB carriers. As a trial, we checked the NIR absorption spectra of the gas-phase C$_{60}^-$ that was obtained at room temperature, and we found that the DIB \textcolor{black}{candidate} $\lambda 10650$ was close to the main absorption band of C$_{60}^-$ at 10655 \r{A}. The detection of weaker subbands in astronomical spectra and in the laboratory spectra of gas-phase C$_{60}^-$ at lower temperatures will be necessary to confirm the identification of this ion. 

\end{enumerate}

It will be of great interest to investigate the correlations among DIBs, particularly the correlations of all DIBs with the C$_{60}^+$ DIBs at $\lambda =$ 9577 and 9632 \r{A}, to assess the contribution of fullerenes and their associated compounds to DIBs. 
Moreover, it will also be important to constrain the sizes and structures of the DIB carriers by fitting the molecular rotational contour to the observed DIB profiles \citep[e.g.,][]{hua15}. The new list of DIBs in the NIR range produced by this study represent a valuable input to for further investigations into DIB carriers.

\begin{acknowledgements}

We are grateful to the staffs of Koyama Astronomical Observatory and La Silla Observatory for their support during our observations. This study is based on observations collected at the European Southern Observatory under ESO program 099.C-0850(B). We thank Dr. Mitsunori Araki for the useful comments and suggestions.

WINERED was developed by the University of Tokyo and the
Laboratory of Infrared High-resolution Spectroscopy (LiH),
Kyoto Sangyo University under the financial support of
Grants-in-Aid, KAKENHI, from JSPS (Nos. 16684001,
20340042, and 21840052) and the MEXT Supported Program
for the Strategic Research Foundation at Private Universities
(Nos. S081061 and S1411028).
This study is financially supported by Grants-in-Aid, KAKENHI, from JSPS (Nos. 16K17669, and 21K13969).
S.H. is supported by Grant-in-Aid for JSPS Fellows Grant No. 13J10504. N.K. is supported by JSPS-DST under the Japan-India Science Cooperative Programs during 2013--2015 and 2016--2018. K.F. is supported by KAKENHI (16H07323) Grant-in-Aid for Research Activity start-up. 

\end{acknowledgements}

\end{document}